\documentclass[]{llncs}
\usepackage{graphicx} 
\usepackage[linesnumbered,ruled,vlined]{algorithm2e}
\usepackage{amsmath, amssymb}
\usepackage{verbatim}
\usepackage[dvipsnames,table]{xcolor}
\usepackage{pst-tree}
\allowdisplaybreaks
\usepackage{float}
\usepackage{tikz}

\newcommand\nodeis[1]{\textcircled{\raisebox{0.75pt}{\tiny{ {\!\!\!#1}}}}}
\newcommand\nodei[1]{\textcircled{\raisebox{-0.9pt}{\small {#1}}}}

\newcommand*\squared[1]{\tikz[baseline=(char.base)]{
            \node[shape=rectangle,draw,inner sep=2pt] (char) {#1};}}         

\newcommand{\mcw}{0.5cm}

\newcommand{\remove}[1]{}
\newcommand{\tcg}{\cellcolor{gray!25} }

\newcommand{\mol}[1]{{\langle #1 \rangle}} 
\newcommand{\calS}{\mbox{${\cal S}$}}
\newcommand{\A}{\mbox{${\cal A}$}}
\newcommand{\B}{\mbox{${\cal B}$}}

\newcommand{\trunc}{\mbox{\normalfont \tt trunc}}
\newcommand{\sm}{\mbox{\normalfont \tt sum}}
\newcommand{\cmap}{\mbox{\normalfont \tt c}}
\newcommand{\yf}{\mbox{\normalfont \tt s}}
\newcommand{\tf}{\mbox{\normalfont \tt t}}

\newcommand{\co}{{\sf c}}
\newcommand{\bo}{{\sf b}}

\title{Old and New Results on Alphabetic Codes\thanks{
  This work was partially supported by project SERICS (PE00000014) under the NRRP MUR program funded by the EU-NGEU.}}
\author{Roberto Bruno \and Roberto De Prisco \and Ugo Vaccaro}
\institute{Dipartimento di Informatica, Universit\`a di Salerno,\\ I-84084 Fisciano (SA), Italy.\\
\email{rbruno@unisa.it, robdep@unisa.it, uvaccaro@unisa.it}}

\date{}
\begin{document}
\maketitle

\begin{abstract}
This comprehensive survey examines the field of alphabetic codes, tracing their development from the 1960s to the present day. We explore classical alphabetic codes and their variants, analyzing their properties and the underlying mathematical and algorithmic principles. The paper covers the fundamental relationship between alphabetic codes and comparison-based search procedures and their applications in data compression, routing, and testing. We review optimal alphabetic code construction algorithms, necessary and sufficient conditions for their existence, and upper bounds on the average code length of optimal alphabetic codes. The survey also discusses variations and generalizations of the classical problem of constructing minimum average length alphabetic codes. By elucidating both classical results and recent findings, this paper aims to serve as a valuable resource for researchers and students, concluding with promising future research directions in this still-active field.
\end{abstract}


\section{Introduction}
Alphabetic codes have been the subject of extensive investigations, both in Information Theory and Computer Science, since the early 1960s.
In this paper, we aim 
to provide a comprehensive survey of the research
field related to alphabetic codes,
tracing the main results from their early inception to their current state-of-the-art. 
We will analyze classical alphabetic codes and their many variants, their properties, 
and the mathematical and algorithmic principles underlying their design.

We will also illustrate 
various applications of alphabetic codes, which span across numerous domains such as search algorithms,
data compression, routing, and testing, to name a few. 

In writing this survey, we intend to provide a valuable resource for researchers and students
alike by offering an 
elucidation  (i.e., not just a narrative about who did what) both  of 
classical results (whose description is not always
easy to dig out), and of recent findings. Finally, we will also illustrate
a few promising future directions for this still fertile field of study.

\section{Structure of the paper}
This survey is organized into several parts. In Section \ref{mot} we describe the various motivations
that led researchers to investigate alphabetic codes. More in particular, in Section \ref{AlpSear}
we illustrate in detail the basic correspondence between alphabetic codes and
comparison-based search procedures. Historically, this correspondence was the first
incentive for the study of alphabetic codes and their properties. In Section \ref{add}
we describe several additional application scenarios where alphabetic codes
play an important role.

In Section \ref{sec:optimal} we review the known algorithms to
construct optimal alphabetic codes (that is, of minimum average length).
We also explain in detail
the structure of the most efficient known algorithms with worked examples.

In Section \ref{sec:conditions} we present
three necessary and sufficient conditions for the existence of alphabetic codes. These conditions represent, in a sense, the generalizations of the classical Kraft condition for the existence of prefix codes.

 In Section \ref{sec:upp} we describe explicit upper bounds on the average length of optimal alphabetic codes.
 Interestingly, these upper bounds are often accompanied by \textit{linear} time algorithms to construct alphabetic codes whose average lengths are within such bounds.

 In Section \ref{vargen} and Section \ref{misc} we survey the numerous results
about variations and generalizations of the classical problem
of constructing alphabetic codes of minimum average length.

We conclude the paper with Section \ref{open}, where we list a few
interesting open problems in the area of alphabetic codes.

\section{Motivations}\label{mot}

In the following subsections, we illustrate the main motivations and applications of alphabetic codes.
\subsection{Alphabetical Codes and Search Procedures}\label{AlpSear}  
Search Theory and the Theory of Variable-Length Codes are strongly linked \cite{AW,aigner,Hass,ka,knuth,KM}.
Indeed, \textit{any} search process that sequentially 
executes suitable tests to identify objects within a given 
search space \textit{inherently} produces a variable-length encoding 
for elements in that space. 
Specifically, one can represent each potential test outcome using a distinct symbol from a finite code alphabet, and by concatenating these (encoded) test outcomes, 
one obtains a legitimate encoding of
any object within the search domain.

More in particular, binary \textit{prefix and alphabetic}\footnote{For 
the sake of brevity, from this point
on a \textit{prefix and alphabetic code} will be simply referred as an  \textit{alphabetic code}.} codes emerge as fundamental combinatorial structures in Search Theory; indeed, alphabetic codes are mathematically
\textit{equivalent} to search procedures that operate via binary comparison queries in totally
ordered sets. 
To explain this equivalence,
we first  introduce 
the formal definition of binary \textit{alphabetic codes}.

\begin{definition}\label{def:alp}
    Let $S=\{s_1, \ldots, s_m\}$  be a set of symbols,
ordered according to a given total order relation $\prec$, that is, for which $s_1\prec \dots \prec s_m$ holds. 
A binary alphabetic code is a
mapping $w:\{s_1, \ldots , s_m\}\mapsto \{0,1\}^+$, 
enjoying the following two properties
\begin{itemize}
    \item the mapping 
 $w:\{s_1, \ldots , s_m\}\mapsto \{0,1\}^+$ is order-preserving,
where the order relation on the set of all
binary strings $\{0,1\}^+$ 
is the standard alphabetical order,
\item no codeword $w(s)$ is prefix
of another $w(s')$, for any $s,s'\in S$, $s\neq s'$.
\end{itemize}
We denote by $C$ the set of
codewords 
$$C=\{w(s): s\in S\}.$$
\end{definition}

We illustrate how alphabetic codes arise from search algorithms with the following
examples.
\begin{example}
Let $S=\{1, 2, \ldots , 8\}$ be the search space. We consider a search algorithm
that attempts to determine an unknown element $x\in S$ by asking queries
of the form ``is $x\leq j$?" for $j=1, 2, \ldots , 8.$ The algorithm (and the
corresponding answers to queries) can be represented by the following binary tree.
Each internal node of the tree corresponds to a query ``is $x\leq j$?", and each
branch emanating from a node corresponds either to the Yes answer
to the node query or to the NO answer.
Each leaf $f$ of the tree corresponds to the (unique) element of $S$ that is consistent
    with the sequence of Yes/No answers (to the node questions)
    from the root of the tree to the leaf $f$.

\bigskip
\begin{figure}[H]
    \centering
    \hspace{-3cm}
    \includegraphics{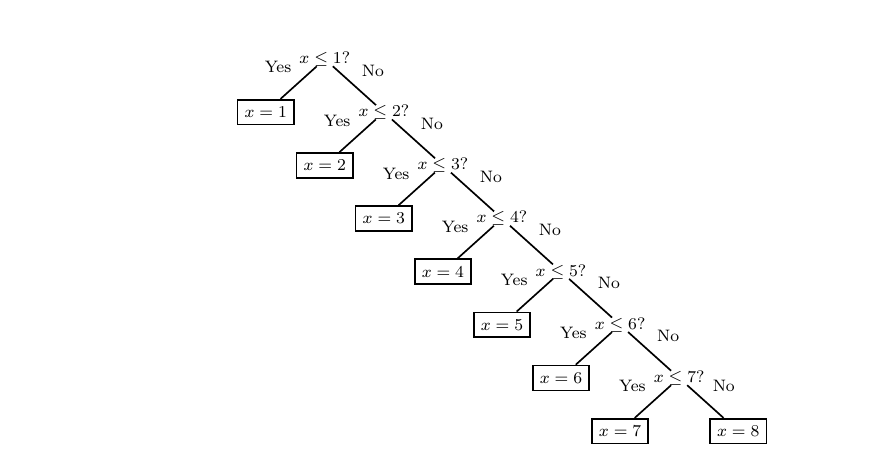}
\end{figure}
\vspace*{1cm}

By encoding the Yes answer to each test with the symbol 0 and the
No answer with 1,
we get a binary coding  $\co:\{1, \ldots , 8\}\mapsto \{0,1\}^+$, namely:
\co(1)=0,  \co(2)=10, 
\co(3)=110, \co(4)=1110, 
\co(5)=11110, \co(6)=111110, 
\co(7)=1111110, 
\co(8)=1111111, 
that is clearly alphabetic.
Note that the length of the $i^{th}$ codeword
corresponds to the number of tests required to determine whether or not 
the unknown element $x$ is equal to $i$, for each $i\in S$.

\textit{Conversely}, if one had the binary alphabetic coding  
$\co:\{1, \ldots , 8\}\mapsto \{0,1\}^+$ defined above, 
it would be easy to design an algorithm $\cal A$ that searches successfully in the space
$\{1, \ldots , 8\}$. More precisely, one could partition the search
space 
$S=\{1, \ldots , 8\}$ in 
$$S_0=\{i\in S : \mbox{ the first bit of } \co(i) \mbox{ is } 0\}$$
 { and }
$$S_1=\{i\in S : \mbox{ the first bit of } \co(i) \mbox{ is } 1\}.$$
Let $m$ be the maximum of the set $S_0$. The first query of the
algorithm $\cal A$ is ``is $x\leq m$?", where $x$ is the unknown element in $S$
we are trying to determine. Since the encoding $\co$ is alphabetic, we know that 
both 
$S_0$ and $S_1$ are made by \textit{consecutive} elements of $S$. 
Therefore, the answer to the query ``is $x\leq m$?" allows one to identify
the first bit of the encoding of the unknown $x$. This way to
proceed can be iterated either in $S_0$ or $S_1$ (according to the query's response) until one gets all bits of the encoding $\co(x)$. 
From the knowledge of $\co(x)$ one gets the value of the unknown element $x$.
\end{example}

\begin{example}
One could use a different algorithm to determine an unknown element $x\in S$.
For example, a binary search that performs, at each step,
the query ``is $x\leq j$?", where $j$ is the middle point of the interval
that contains $x$. In this case, the tree representing the algorithm is: 
\bigskip

\begin{figure}[H]
    \centering
    \hspace{-3cm}
    \includegraphics{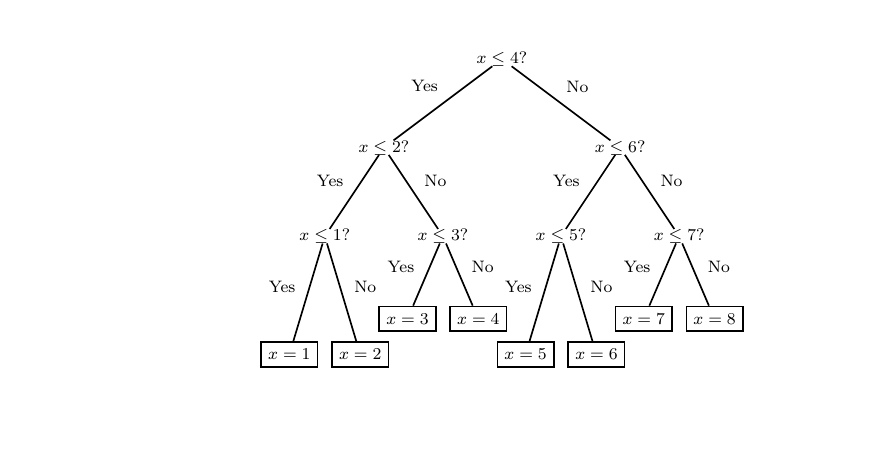}
\end{figure}

Again, by encoding the  Yes answer for each test with the symbol 0 and the No answer with 1 we get the
(different) encoding
of $1, \ldots , 8$, given by: 
\bo(1)=000,
 \bo(2)=001,
\bo(3)=010,
\bo(4)=011,
\bo(5)=100,
\bo(6)=101,
\bo(7)=110,
\bo(8)=111. 
Also in this case one can see that the 
obtained encoding  $\bo(\cdot)$ is order-preserving,
and therefore alphabetic. As before, from the encoding  $\bo(\cdot)$ one 
can easily design an algorithm  that successfully searches in the space
$\{1, \ldots , 8\}$. The idea is always the same:
The search
space 
$S=\{1, \ldots , 8\}$ can be partitioned in 
$$S_0=\{i\in S : \mbox{ the first bit of } \bo(i) \mbox{ is } 0\}$$  
and
$$S_1=\{i\in S : \mbox{ the first bit of } \bo(i) \mbox{ is } 1\}.$$
Since the encoding $\bo$
is alphabetic, we know that 
both 
$S_0$ and $S_1$ are made by \textit{consecutive} elements of $S$ (in our case,
$S_0=\{1, 2, 3, 4\}$ and $S_1=\{5, 6, 7, 8\}$. 
Let $m$ be the maximum of the set $S_0$. The first query of the
algorithm $\cal A$ is ``is $x\leq m$?", where $x$ is the unknown element in $S$
we are trying to determine. According to the answer to the query, the algorithm $\cal A$
will recursively iterate in $S_0$ or in $S_1$.
\end{example}

\medskip

In general, it holds the following basic result.
\begin{theorem}[\cite{AW,aigner}]
    Let $S=\{s_1, \ldots, s_m\}$  be a set of elements,
ordered according to a given total order relation $\prec$, that is, for which it holds that
$s_1\prec \dots \prec s_m$. \emph{Any} algorithm $\cal A$ that successfully determines 
the value of an arbitrary unknown $x\in S$, by means of the execution of tests 
of the type ``is $x\prec s$?", for given $s\in S$, gives rises to a 
{prefix and alphabetic} binary encoding 
of the elements of $S$.

Conversely, from \emph{any} prefix and alphabetic binary encoding
of the elements of $S$ one can construct
an algorithm $\cal A$ that successfully determines
the value of an arbitrary unknown $x\in S$, by means of the execution of tests
of the type ``is $x\prec s$?".
\end{theorem}

In the rest of this paper, we will use the correspondence above described
between trees and codes, in the sense that we will freely switch between 
the terminology of codes and trees,  according
to which is more suitable for the scenario we will be considering.

\subsection{Additional applications of alphabetic codes}\label{add}
In addition to search problems, alphabetic codes arise in several other circumstances.
They have been used in \cite{gupta2000near} to provide efficient algorithms
for the routing lookup problem. Indeed, standard \textit{classless interdomain routing} requires that a router performs a ``longest prefix match”
to determine the next hop of a packet. 
Therefore,  given a packet,
the lookup operation consists of finding the longest prefix in the
routing table that matches the first few bits of the destination
address of the packet. In the paper \cite{gupta2000near} the authors show
how alphabetic codes allow one to speed up the above-described operation.
Subsequently, the author of \cite{Nagaraj2011482} somewhat improved the
analysis contained in the article \cite{gupta2000near}, always using alphabetic codes
as a basic tool.

In the paper 
\cite{Vaishnav} the authors apply (variants of) alphabetic trees to
problems arising in efficient VLSI design. More specifically,
they consider the problem of \textit{fan-out optimization}, whereby
one tries to design logical circuits with bounded fan-out.
Interestingly, the authors of \cite{Vaishnav} show that, 
after appropriate technology-independent optimization, 
the fan-out optimization problem essentially becomes a tree optimization problem;
subsequently, 
they develop suitable alphabetic tree generation and optimization algorithms, and 
apply them to the fan-out optimization problem. 

The papers \cite{L+,pattipati1990application,Yeung} applied ideas, techniques and
results about 
alphabetic codes to the problem of
designing efficient algorithms for noiseless fault diagnosis. Similarly, the paper \cite{garey1972optimal} considered the application of alphabetic codes to binary identification procedures that go from machine fault location to medical diagnosis
and more. In the paper \cite{Graefe} the author discussed the use of alphabetic codes 
for order-preserving data compression in the implementation of database systems, 
to the purpose of saving  space and bandwidth at all levels of the memory
hierarchy. 
The paper \cite{Gagie} exploited the properties of alphabetic codes to design
efficient algorithms for compression of probability distributions.   
Finally, the paper \cite{Ar}
applied alphabetic code to the problem of the efficient design of 
encryption algorithms.

Alphabetic codes are also useful for the implementation of arithmetic coding: Because
binary arithmetic coding is much faster than other types of arithmetic
coding, a decision tree (representing an alphabetic code) can 
be used to reduce an infinite alphabet source into
a binary source for fast arithmetic coding, as done in \cite{Mar}. In addition, 
the basic order preservation property of alphabetic codes is
necessary for the ordered representation of
rational numbers as integers in continued fractions (e.g., see \cite{Mat,Yo}).

Moreover, in the research paper \cite{pe+} 
the authors used alphabetic codes as a  tool for the construction
of variable-length unidirectional error-detecting codes with few check symbols.

Alphabetic codes were also used in computational geometry; more precisely, in \cite{Pr} 
they have been used for efficiently locating a point on a line
when the query point does
not coincide with any of the points dividing the line.

We conclude this section by mentioning that alphabetic
codes are strictly related to \textit{binary search trees} \cite{Nagaraj}, a
very important data structure widely used in many computer science applications.
In a sense, binary search trees constitute a generalization of
alphabetic codes, in that search algorithms that give rise to 
binary search trees operate by comparison \textit{and} equality tests;
moreover, binary search trees 
take into account successful \textit{and unsuccessful} 
searches, while alphabetic codes can be considered particular
case of search trees in which successful searches have zero probability of occurrences.
For a nice survey on binary search trees and their many applications, we refer the
reader to the paper \cite{Nagaraj}.

\section{Algorithms for constructing  optimal alphabetic codes}\label{sec:optimal}

Recall that we denote with $S=\{s_1, \ldots, s_n\}$ the set of symbols and 
that, over such a set, we have  
a total order relation $\prec$, for which it holds
$s_1\prec \dots \prec s_n$. 
We assume that the set $S$ is endowed with a probability
distribution $P=\mol{p_1, \ldots , p_n}$, that is, $p_i$ is
the probability of symbol $s_i$, for $i=1, \ldots , n.$ To emphasize that we are dealing
with ordered lists, we use the notation $\mol{\cdot}$.
Given an order-preserving mapping
$w:s\in \{s_1, \ldots , s_n\}\mapsto w(s)\in \{0,1\}^+$, 
we denote the average code length of the alphabetic code $C=\{w(s): s\in S\}$ by
\begin{equation}\label{E[C]}
    \mathbb{E}[C]= \sum_{i=1}^m p_i\,\ell_i, 
\end{equation}
where $\ell_i$ is the length of $w(s_i)$.
The basic problem is to find efficient algorithms
to construct alphabetic codes for which the parameter (\ref{E[C]}) is \textit{minimum}.
We recall that the minimum possible value of (\ref{E[C]}) is lower bounded
by the Shannon entropy $H(P)=-\sum_ip_i\log p_i$ of $P$.

In their classic paper \cite{gilbert1959variable}, 
Gilbert and Moore designed a
dynamic programming algorithm, of time
complexity $O(n^3)$, for the construction of optimal alphabetic codes, that is, of
minimum average length. 
Subsequently, Knuth \cite{knuthp} gave an improved $O(n^2)$ algorithm. 
Hu and Tucker   provided
 an algorithm of time complexity $O(n\log n)$, with a fairly complicated correctness proof that was later slightly simplified by Hu \cite{Hu_proof}.
Garsia and Wachs \cite{Garsia_al} gave a similar algorithm, which has been shown to be equivalent to the Hu-Tucker algorithm \cite{Nagaraj}.
Kingston has provided a simpler analysis of the Garsia-Wachs algorithm
in the paper \cite{KINGSTON}. Similarly, Karpinski, Larmore, and Rytter \cite{Karp} gave new correctness proofs for both the Garsia-Wachs algorithm and the Hu-Tucker algorithm. 
Finally,  in  \cite{belal2002building} 
Belal \textit{et al.} gave a different algorithm and claimed  
that it produces  optimal 
alphabetic codes.


In the rest of this section, we provide a description of the algorithms and an
example to aid in the explanation. 
For the example, we will use $n=11$ and the following probability distribution
$$P = \mol{0.24,0.12,0.09, 0.08, 0.04, 0.02, 0.03, 0.06, 0.14, 0.11, 0.07}.$$

\subsection{Gilbert and Moore's algorithm}

The Gilbert-Moore algorithm for the construction of optimal alphabetic codes is a dynamic programming algorithm.
We can define the subproblems $S(i,j)$ as the construction of an optimal (sub)tree for the symbols $s_i,\ldots,s_j$, with $1 \le i\le j\le n$.
The complete problem is that of finding an optimal alphabetic code/tree for $S(1,n)$.
The borderline cases are the subproblems with $i=j$ for which the cost is $C(i,i)=0$, since there is nothing to encode,
and the subproblems with $j=i+1$, that is, those with only two consecutive symbols, for which the optimal code assigns 
the two codewords 0 and 1 to the two symbols, or, in other words, for which the optimal tree is a root with two children that are leaves.
The cost is $C(i,i+1)=p_i+p_{i+1}$, for $i=1,2,\ldots,n-1$. Table~\ref{tab:gm:dynpro:init} shows these costs for $n=11$.

\begin{table}[h]
\centering{
\resizebox{1\textwidth}{!}{
\begin{tabular}{|c|c|c|c|c|c|c|c|c|c|c|c|}
\hline
$i\backslash j$ & 1 & 2 & 3 & 4 & 5 & 6 & 7 & 8 & 9 & 10 & 11\\ 
\hline
\hline
1 & \makebox[1cm]{0} & $p_1+p_2$ &  &  &  &  &  &  &  &  & \\ 
\hline
2 &\tcg& 0 & $p_2+p_3$ &  &  &  &  &  &  &  & \\ 
\hline
3 &\tcg&\tcg& 0 & $p_3+p_4$ &  &  &  &  &  &  & \\ 
\hline
4 &\tcg&\tcg&\tcg& 0 & $p_4+p_5$ &  &  &  &  &  & \\ 
\hline
5 &\tcg&\tcg&\tcg&\tcg& 0 & $p_5+p_6$  &  &  &  &  & \\ 
\hline
6 &\tcg&\tcg&\tcg&\tcg&\tcg& 0 & $p_6+p_7$  &  &  &  &  \\ 
\hline
7 &\tcg&\tcg&\tcg&\tcg&\tcg&\tcg& 0 & $p_7+p_8$ &  &  & \\ 
\hline
8 &\tcg&\tcg&\tcg&\tcg&\tcg&\tcg&\tcg& 0 &  $p_8+p_9$ &  & \\ 
\hline
9 &\tcg&\tcg&\tcg&\tcg&\tcg&\tcg&\tcg&\tcg& 0 & $p_9+p_{10}$ & \\ 
\hline
10 &\tcg&\tcg&\tcg&\tcg&\tcg&\tcg&\tcg&\tcg& \tcg& 0 & $p_{10}+p_{11}$  \\ 
\hline
11 &\tcg&\tcg&\tcg&\tcg&\tcg&\tcg&\tcg&\tcg&\tcg&\tcg& 0   \\ 
\hline 
\end{tabular}
}
}
\caption{Initial matrix for the dynamic programming algorithm ($n=11$).}\label{tab:gm:dynpro:init}
\end{table}

Then the optimal tree for the subproblem $S(i,j)$ can be found by checking all the possible ways of splitting the sequence
of symbols $s_i,\ldots,s_j$ into a left and a right subtree with at least one node in each subtree. There are exactly $j-i$ ways
to perform such a split, namely $s_i,\ldots,s_k$ on the left and $s_{k+1},\ldots,s_j$ on the right, for $k=i,i+1,\ldots,j-1$.
The cost of the tree produced by the split for a given value $k$ is
$$C(i,j)=\sum_{s=i}^j p_s + C(i,k) + C(k+1,j).$$

For example, to compute $C(1,3)$ we consider the two possible splits $1:2..3$ and $1..2:3$. The first one
has cost $(p_1+p_2+p_3)+C(1,1)+C(2,3)= p_1+2p_2+2p_3$ and the second one
has cost $(p_1+p_2+p_3)+C(1,2)+C(3,3)= 2p_1+2p_2+p_3$. The minimum cost determines the optimal cost
for the subproblem $S(1,3)$. It is easy to fill the cost table with an $O(n^3)$ algorithm, 
and some easy bookkeeping allows one to build the optimal tree/code, as illustrated in 
Algorithm~\ref{alg:gilbert-moor:optimal}.

\begin{algorithm}[H]\label{alg:gilbert-moor:optimal}
\caption{Gilbert-Moore algorithm}
\KwIn{Symbols $S=\{s_1,\dots,s_n\}$ and $P=\mol{p_1,\dots,p_n}$ the associated probability distribution.}
$C(1,1)=0$

\For{$i \gets 2$ \textbf{to} $n$} {
    $C(i,i)=0$
    
    $C(i-1,i)=p_{i-1}+p_{i}$
}

\For{$s \gets 2$ \textbf{to} $n-1$} { 
    \For{$i \gets 1$ \textbf{to} $n-s$} {

         $j=i+s$ \tcp*{Proceed by diagonals}
    
         $min=\infty$
         
         $minindex = -1$
         
         \For{$k \gets i$ \textbf{to} $j-1$} {
            \If {$C(i,k)+C(k+1,j)<min$} {
                $min=C(i,k)+C(k+1,j)$

                $minindex=k$
                }
          }
          $C(i,j)=\sum_{k=i}^j p_k+min$
          
          $R(i,j)=minindex$
    }
}
\KwOut{$C(1,n)$ and $R$}
\end{algorithm}

\begin{table}
\resizebox{1.0\textwidth}{!}{
\begin{tabular}{|c|c|c|c|c|c|c|c|c|c|c|c|}
\hline
$i\backslash j$ & \makebox[\mcw]{1} & \makebox[\mcw]{2}  & \makebox[\mcw]{3}  & \makebox[\mcw]{4}  & \makebox[\mcw]{5}  & \makebox[\mcw]{6}  & \makebox[\mcw]{7}  & \makebox[\mcw]{8}  & \makebox[\mcw]{9}  & \makebox[\mcw]{10}  & \makebox[\mcw]{11} \\ 
\hline
\hline
 1 &    0&  36&  66&  99& 123& 135& 152& 182& 236& 283& 322\\
\hline
 2 & \tcg&   0&  21&  46&  66&  76&  90& 116& 162& 209& 242\\
\hline
 3 & \tcg&\tcg&   0&  17&  33&  43&  57&  78& 120& 160& 192\\
\hline
 4 & \tcg&\tcg&\tcg&   0&  12&  20&  31&  51&  88& 124& 156\\
\hline
 5 & \tcg&\tcg&\tcg&\tcg&   0&   6&  14&  29&  58&  94& 123\\
\hline
 6 & \tcg&\tcg&\tcg&\tcg&\tcg&   0&   5&  16&  41&  77& 102\\
\hline
 7 & \tcg&\tcg&\tcg&\tcg&\tcg&\tcg&   0&   9&  32&  66&  91\\
\hline
 8 & \tcg&\tcg&\tcg&\tcg&\tcg&\tcg&\tcg&   0&  20&  51&  76\\
\hline
 9 & \tcg&\tcg&\tcg&\tcg&\tcg&\tcg&\tcg&\tcg&   0&  25&  50\\
\hline
10 & \tcg&\tcg&\tcg&\tcg&\tcg&\tcg&\tcg&\tcg&\tcg&   0&  18\\
\hline
11 & \tcg&\tcg&\tcg&\tcg&\tcg&\tcg&\tcg&\tcg&\tcg&\tcg&   0\\
\hline
\end{tabular}
\hspace{1truecm}
\begin{tabular}{|c|c|c|c|c|c|c|c|c|c|c|c|}
\hline
$i\backslash j$ & \makebox[\mcw]{1} & \makebox[\mcw]{2}  & \makebox[\mcw]{3}  & \makebox[\mcw]{4}  & \makebox[\mcw]{5}  & \makebox[\mcw]{6}  & \makebox[\mcw]{7}  & \makebox[\mcw]{8}  & \makebox[\mcw]{9}  & \makebox[\mcw]{10}  & \makebox[\mcw]{11} \\ 
\hline
\hline
 1 & \tcg&   1&   1&   1&   1&   1&   1&   2&   3&   3&   3\\
\hline
 2 & \tcg&\tcg&   2&   2&   2&   3&   3&   3&   4&   4&   7\\
\hline
 3 & \tcg&\tcg&\tcg&   3&   3&   3&   3&   4&   5&   8&   8\\
\hline
 4 & \tcg&\tcg&\tcg&\tcg&   4&   4&   4&   5&   7&   8&   8\\
\hline
 5 & \tcg&\tcg&\tcg&\tcg&\tcg&   5&   5&   7&   8&   8&   9\\
\hline
 6 & \tcg&\tcg&\tcg&\tcg&\tcg&\tcg&   6&   7&   8&   8&   9\\
\hline
 7 & \tcg&\tcg&\tcg&\tcg&\tcg&\tcg&\tcg&   7&   8&   9&   9\\
\hline
 8 & \tcg&\tcg&\tcg&\tcg&\tcg&\tcg&\tcg&\tcg&   8&   9&   9\\
\hline
 9 & \tcg&\tcg&\tcg&\tcg&\tcg&\tcg&\tcg&\tcg&\tcg&   9&   9\\
\hline
10 & \tcg&\tcg&\tcg&\tcg&\tcg&\tcg&\tcg&\tcg&\tcg&\tcg&10\\
\hline
11 & \tcg&\tcg&\tcg&\tcg&\tcg&\tcg&\tcg&\tcg&\tcg&\tcg&\tcg\\
\hline
\end{tabular}
}
\caption{Costs (left) and roots indexes (right) matrices. Costs are multiplied by 100.}\label{tab:dp:ex}
\end{table}

Table~\ref{tab:dp:ex} shows the values obtained for the probability distribution $P$ (in the table
the values of the costs are multiplied by 100).
The cost of an optimal tree is $3.22$. The roots indexes matrix allows to build the
tree; for example, $R(1,11)=3$ means that the first split of the optimal tree described
by this matrix is $1..3:4..11$.

\subsection{Knuth's algorithm}

Knuth proved that the search of the root of the optimal subtree, which, for each element $(i,j)$ of the matrix, takes $j-i$ iterations (the {\bf for} loop at line 9 in Algorithm~\ref{alg:gilbert-moor:optimal}),
can be restricted to a smaller interval that depends on the roots of the smaller subtrees. Namely, Knuth proved that the root $R(i,j)$ of an
optimal tree for the subproblem $S(i,j)$, can be found between the indexes $R(i,j-1)$ and $R(i+1,j)$. Thus instead of searching from $i$ to $j-1$, it is
sufficient to search from $R(i,j-1)$ through $R(i+1,j)$.
This means that the for loop at line 9 can be changed to

\medskip

\centerline{{\bf for} $k=R(i,j-1)$ {\bf to} $R(i+1,j)$ {\bf do}}

\medskip

\noindent
with savings on the total execution time that lowers the time complexity of the algorithm to $O(n^2)$.

To better understand this saving we report in Table~\ref{tab:knuth:roots} the search intervals for the 
Gilbert-Moore dynamic programming algorithm
and the search intervals of Knuth's improvement on the input of the previous example.
For the Gilbert-Moore algorithm, the size of the interval is fixed (because it depends only on the
indexes $i$ and $j$ and in each iteration the size grows by $1$), while for the Knuth's algorithm it
depends on the input that determines the roots of the subtrees, and it is smaller.

\begin{table}
\resizebox{1.0\textwidth}{!}{
\begin{tabular}{|c|c|c|c|c|c|c|c|c|c|c|c|}
\hline
$i\backslash j$ & \makebox[\mcw]{1} & \makebox[\mcw]{2}  & \makebox[\mcw]{3}  & \makebox[\mcw]{4}  & \makebox[\mcw]{5}  & \makebox[\mcw]{6}  & \makebox[\mcw]{7}  & \makebox[\mcw]{8}  & \makebox[\mcw]{9}  & \makebox[\mcw]{10}  & \makebox[\mcw]{11} \\ 
\hline
\hline
 1 & \tcg&   1&   1-2&   1-3&   1-4&   1-5&   1-6&   1-7&   1-8&   1-9&   1-10\\
\hline
 2 & \tcg&\tcg&   2   &  2-3&   2-4&   2-5&  2-6&   2-7&   2-8&   2-9& 2-10\\
\hline
 3 & \tcg&\tcg&\tcg&   3&   3-4&   3-5&   3-6&   3-7&   3-8&   3-9&   3-10\\
\hline
 4 & \tcg&\tcg&\tcg&\tcg&   4&   4-5&   4-6&   4-7&   4-8&   4-9&   4-10\\
\hline
 5 & \tcg&\tcg&\tcg&\tcg&\tcg&   5&   5-6&   5-7&   5-8&   5-9&   5-10\\
\hline
 6 & \tcg&\tcg&\tcg&\tcg&\tcg&\tcg&   6&   6-7&   6-8&   6-9&   6-10\\
\hline
 7 & \tcg&\tcg&\tcg&\tcg&\tcg&\tcg&\tcg&   7&   7-8&   7-9&   7-10\\
\hline
 8 & \tcg&\tcg&\tcg&\tcg&\tcg&\tcg&\tcg&\tcg&   8&   8-9&   8-10\\
\hline
 9 & \tcg&\tcg&\tcg&\tcg&\tcg&\tcg&\tcg&\tcg&\tcg&   9&   9-10\\
\hline
10 & \tcg&\tcg&\tcg&\tcg&\tcg&\tcg&\tcg&\tcg&\tcg&\tcg&10\\
\hline
11 & \tcg&\tcg&\tcg&\tcg&\tcg&\tcg&\tcg&\tcg&\tcg&\tcg&\tcg\\
\hline
\end{tabular}
\hspace{1truecm}
\begin{tabular}{|c|c|c|c|c|c|c|c|c|c|c|c|}
\hline
$i\backslash j$ & \makebox[\mcw]{1} & \makebox[\mcw]{2}  & \makebox[\mcw]{3}  & \makebox[\mcw]{4}  & \makebox[\mcw]{5}  & \makebox[\mcw]{6}  & \makebox[\mcw]{7}  & \makebox[\mcw]{8}  & \makebox[\mcw]{9}  & \makebox[\mcw]{10}  & \makebox[\mcw]{11} \\ 
\hline
\hline
 1 & \tcg&   1&   1-2&   1-2&   1-2&   1-3&   1-3&   1-3&   2-4&   2-4&   3-7\\
\hline
 2 & \tcg&\tcg&   2   &  2-3&   2-3&   2-3&  3-3&   3-4&   3-5&   4-8& 4-8\\
\hline
 3 & \tcg&\tcg&\tcg&   3&   3-4&   3-4&   3-4&   3-5&   4-7&   5-8&   8-8\\
\hline
 4 & \tcg&\tcg&\tcg&\tcg&   4&   4-5&   4-5&   4-7&   5-8&   7-8&   8-9\\
\hline
 5 & \tcg&\tcg&\tcg&\tcg&\tcg&   5&   5-6&   5-7&   7-8&   8-8&   8-9\\
\hline
 6 & \tcg&\tcg&\tcg&\tcg&\tcg&\tcg&   6&   6-7&   7-8&   8-9&  8-9\\
\hline
 7 & \tcg&\tcg&\tcg&\tcg&\tcg&\tcg&\tcg&   7&   7-8&   8-9&   9-9\\
\hline
 8 & \tcg&\tcg&\tcg&\tcg&\tcg&\tcg&\tcg&\tcg&   8&   8-9&  9-9\\
\hline
 9 & \tcg&\tcg&\tcg&\tcg&\tcg&\tcg&\tcg&\tcg&\tcg&   9&   9-10\\
\hline
10 & \tcg&\tcg&\tcg&\tcg&\tcg&\tcg&\tcg&\tcg&\tcg&\tcg&10\\
\hline
11 & \tcg&\tcg&\tcg&\tcg&\tcg&\tcg&\tcg&\tcg&\tcg&\tcg&\tcg\\
\hline
\end{tabular}
}
\caption{Search intervals of root indexes of Gilbert and Moore's algorithm (left) and of Knuth's improvement on the input of the previous example (right).}\label{tab:knuth:roots}
\end{table}

\subsection{Hu and Tucker's algorithm}

The Hu-Tucker algorithm builds an optimal alphabetic tree by first constructing a tree $T'$ that does not preserve the order and then, by exploiting the structure of the obtained tree, it builds a new tree $T$ that maintains the original order. Let us start by describing the first phase in which the tree $T'$ is built. 
The construction of $T'$ is somewhat similar to the construction of a Huffman tree for which the two
smallest probabilities are repeatedly merged together. However there are two crucial differences.
First, two probabilities can be merged together only if, in the ordered list
maintained during the construction, there are no nodes that correspond to single symbols in between the two
probabilities;
this constraint is vacuously satisfied for consecutive probabilities. We will say that two
probabilities are {\em joinable} if they satisfy the constraint. Second, since we are dealing with an ordered list, it
is important to specify where the new element is placed; when joining two probabilities,
the resulting probability takes the place of the ``left'' one, while the ``right'' one gets deleted.
Finally, in case of a tie, that is, when there are two pairs of joinable nodes with the smallest possible sum,
the algorithm always chooses the leftmost one.

\begin{figure}[h]
\centering
\includegraphics[width=0.75\textwidth]{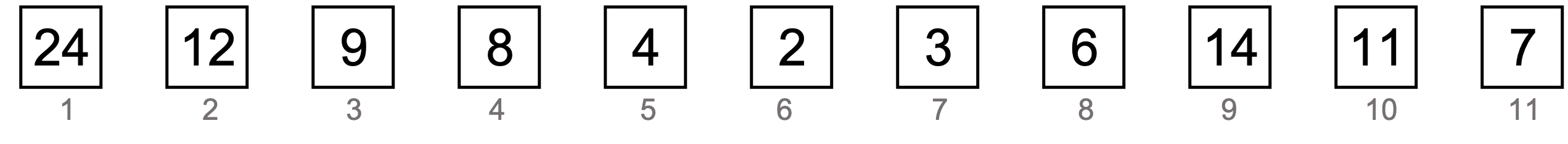} 
\caption{Initial forest. Probabilities are multiplied by 100 to ease the drawing and the reading.}
\label{fig:initial:list}
\end{figure}

The construction starts by creating
a leaf of the tree for each symbol/probability, so initially we have a forest with $n$ trees consisting
of just one node. To clarify the construction we will use an example alongside the description of the steps.
Figure~\ref{fig:initial:list} shows the initial forest for the probability distribution $P$. 
To easily visualize the constraint that makes two nodes joinable, nodes that correspond to leaves are depicted
as squares and internal nodes as circles: two nodes are joinable if in the list there are no squares in
between them.

\begin{figure}[h]
\centering
\includegraphics[width=0.65\textwidth]{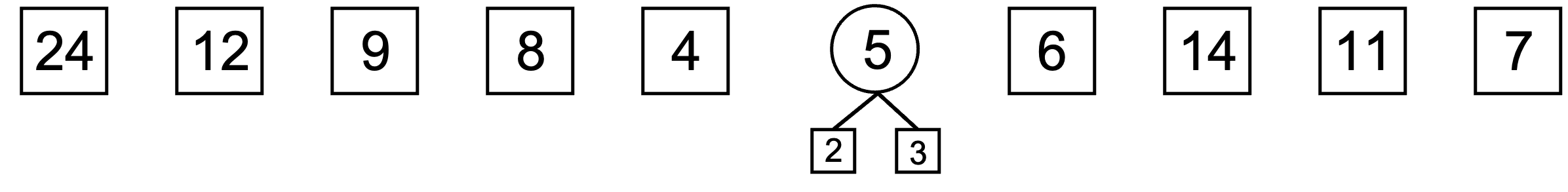} 
\caption{List of nodes after step 1 of the Hu-Tucker algorithm.}
\label{fig:ht:step1}
\end{figure}

In the first step, all pairs of consecutive leaves, and only those pairs, are joinable. Thus the nodes that will be joined
are \squared{2} and \squared{3} because they give the smallest sum. Figure~\ref{fig:ht:step1} shows the resulting forest. 
In the figures we show both
the ordered list of nodes that have to be joined (the top row), and the nodes that have already been joined, by attaching to each node in the list the subtree created by the
joining process.

\begin{figure}[h]
\centering
\includegraphics[width=0.65\textwidth]{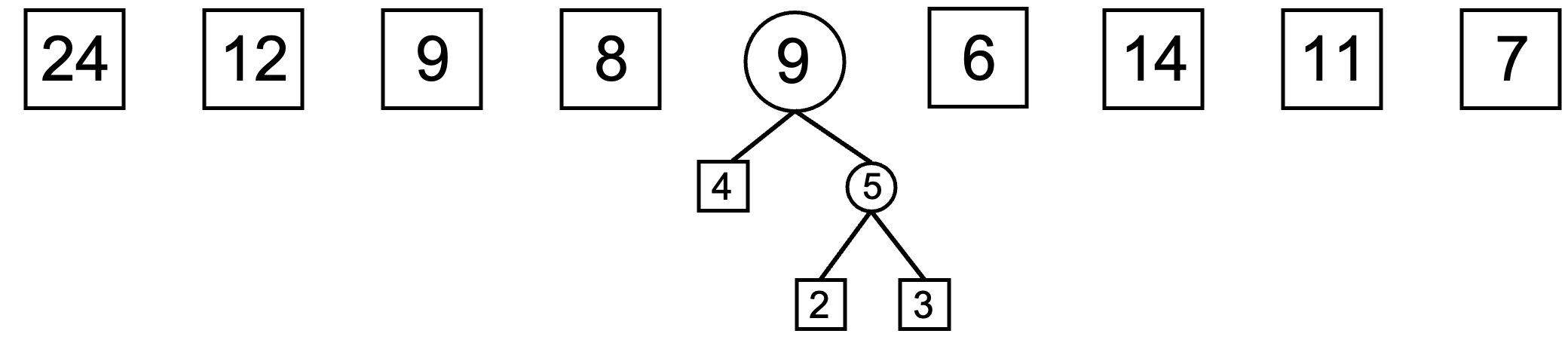} 
\caption{List of nodes after step 2 of the Hu-Tucker algorithm.}
\label{fig:ht:step2}
\end{figure}

Node \nodei{5}, takes the place of node \squared{2}, while node \squared{3} is deleted from the sequence
of nodes that have to be joined. Recall that the newly created node takes the place of the leftmost node among the joined nodes
(in this case it does not make a difference since the two joined nodes are adjacent).
The list of nodes that have to be joined is now 
$\mol{\squared{24},\squared{12},\squared{9},\squared{8},\squared{4},\nodei{5},\squared{6},\squared{14}, \squared{11},\squared{7}}$.
In the second step, the pairs of nodes that are joinable are all consecutive pairs of nodes, but now
 nodes \squared{4} and \squared{6} also are joinable since in between them there are no squares (leaves). 
The smallest sum is given by the pair \squared{4} and \nodei{5} whose joining gives the forest shown in Figure~\ref{fig:ht:step2}. 

The newly created node
takes the position of the node \squared{4} in the list of remaining nodes.

\begin{figure}[h]
\centering
\includegraphics[width=0.6\textwidth]{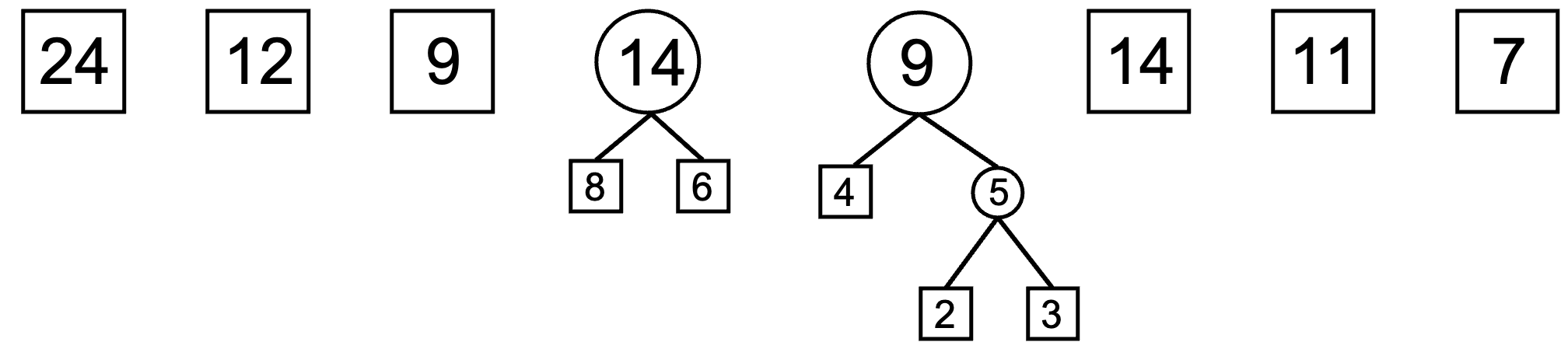} 
\caption{List of nodes after step 3 of the Hu-Tucker algorithm.}
\label{fig:ht:step3}
\end{figure}

For the next step, the list of nodes is
$\mol{\squared{24},\squared{12},\squared{9},\squared{8},\nodei{9},\squared{6},\squared{14}, \squared{11},\squared{7}}$
and thus the joinable nodes are all the consecutive pairs of nodes and the pair \squared{8} and \squared{6}.
And this last pair is the one that gives the smallest sum. Their joining produces
the forest shown in Figure~\ref{fig:ht:step3}, with the new node taking the place of node \squared{8}. 
Notice that this step causes the
order of the leaves to be disrupted, as node \squared{6} has moved to the left of node \squared{4}.

\begin{figure}[h]
\centering
\includegraphics[width=0.6\textwidth]{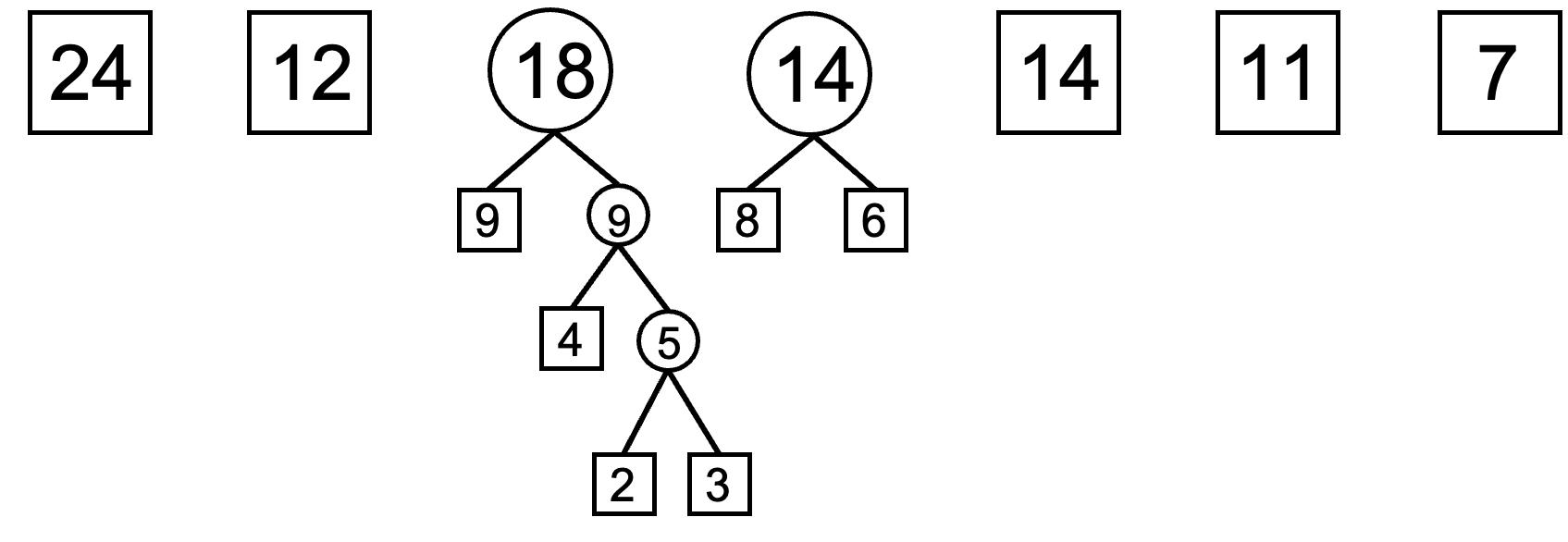} 
\caption{List of nodes after step 4 of the Hu-Tucker algorithm.}
\label{fig:ht:step4}
\end{figure}

In step 4, the joinable nodes are all the consecutive pairs of nodes and the pairs
\squared{9} and \nodeis{9},
\squared{9} and \squared{14} and,
\nodeis{14} and \squared{14}.
In this case, we have that the smallest sum is 18 and is achieved by \squared{9} and \nodeis{9} and by \squared{11} and \squared{7}. The algorithm chooses the leftmost pair, which is \squared{9} and \nodeis{9}, producing the forest shown in Figure~\ref{fig:ht:step4}. 

\begin{figure}[h]
\centering
\includegraphics[width=0.5\textwidth]{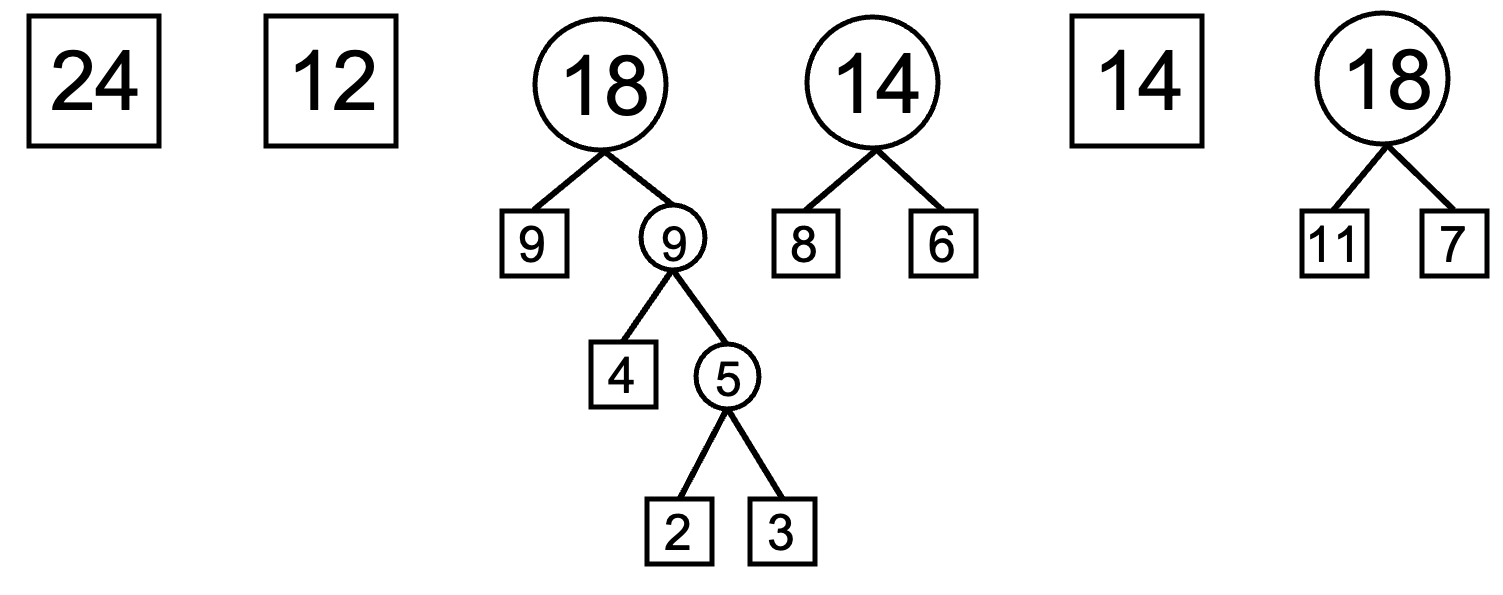} 
\caption{List of nodes after step 5 of the Hu-Tucker algorithm.}
\label{fig:ht:step5}
\end{figure}

For the next step, the set of joinable pairs is again
all the consecutive pairs of nodes and the pairs
\squared{12} and \nodeis{14},
\squared{12} and \squared{14} and,
\nodeis{18} and \squared{14}.
The smallest sum is given by the nodes \squared{11} and \squared{7} and their joining produces the forest shown in Figure~\ref{fig:ht:step5}. 

\begin{figure}[H]
\centering
\includegraphics[width=0.45\textwidth]{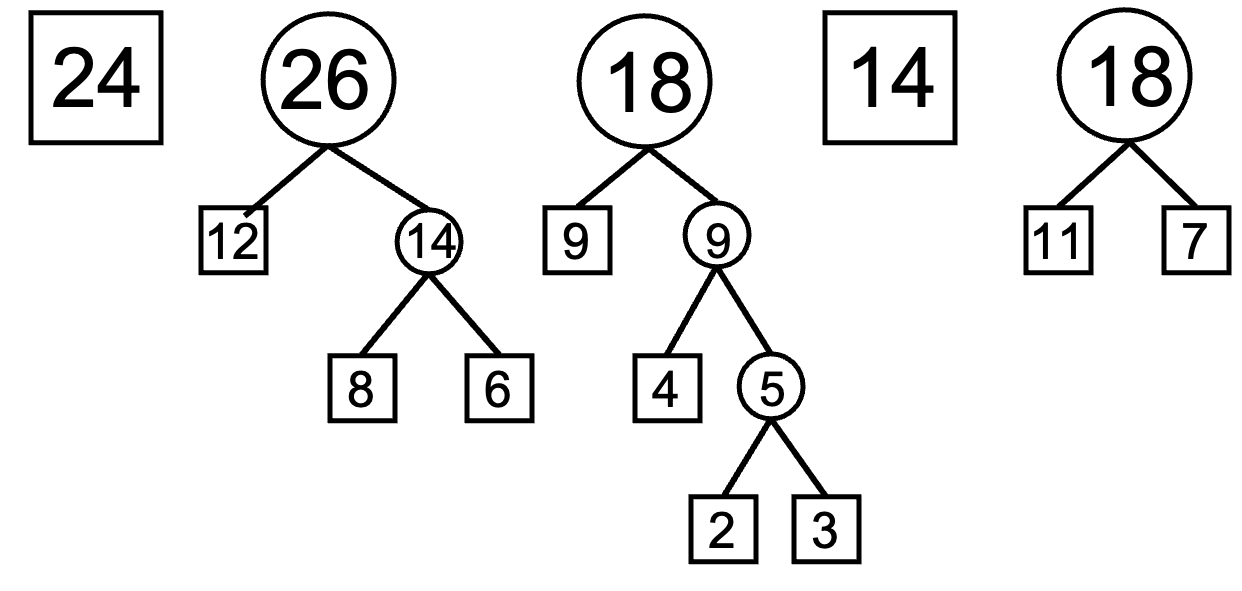} 
\caption{List of nodes after step 6 of the Hu-Tucker algorithm.}
\label{fig:ht:step6}
\end{figure}

For step 6, the joinable pairs are all consecutive nodes, and the pairs
\squared{12} and \nodeis{14},
\squared{12} and \squared{14}, and 
\nodeis{18} and \squared{14}. And the leftmost
smallest sum is obtained by joining the two nodes
\squared{12} and \nodeis{14}, resulting in the forest shown in Figure~\ref{fig:ht:step6}. 

\begin{figure}[H]
\centering
\includegraphics[width=0.45\textwidth]{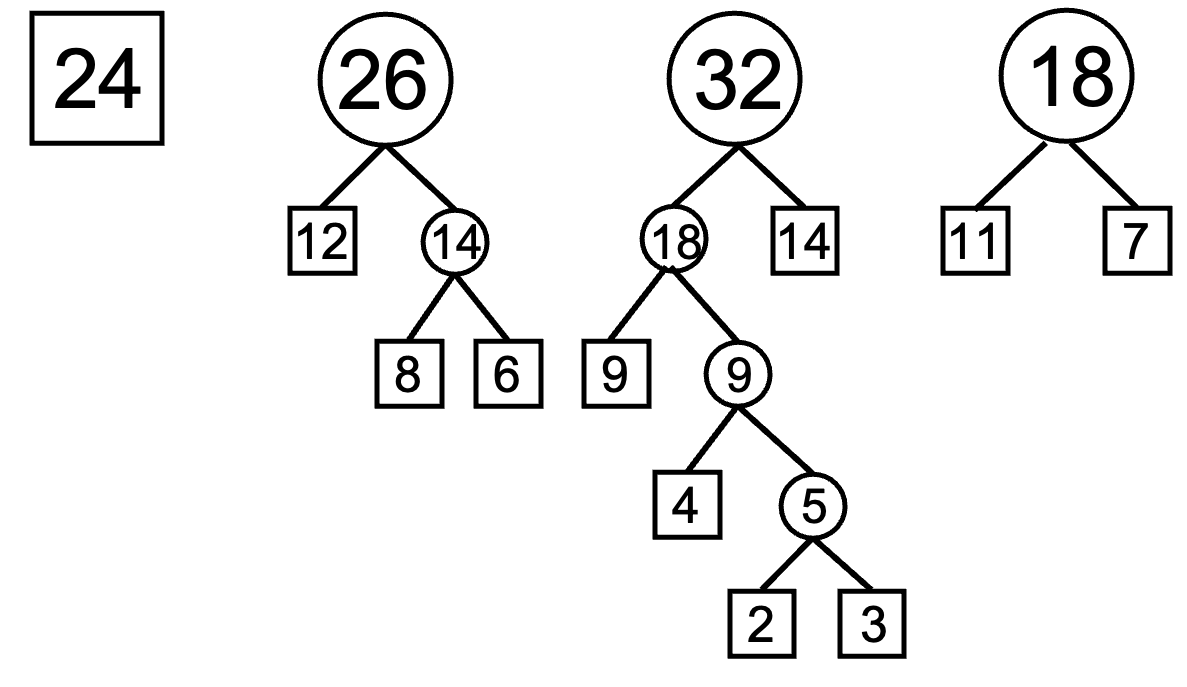} 
\caption{List of nodes after step 7 of the Hu-Tucker algorithm.}
\label{fig:ht:step7}
\end{figure}

For step 7, the joinable pairs are all consecutive nodes, and the pairs
\squared{24} and \nodeis{18},
\squared{24} and \squared{14}, and 
\nodeis{26} and \squared{14}. And the leftmost
smallest sum is obtained by joining the two nodes
\nodeis{18} and \squared{14}, resulting in the forest shown in Figure~\ref{fig:ht:step7}. 

\begin{figure}[h]
\centering
\includegraphics[width=0.6\textwidth]{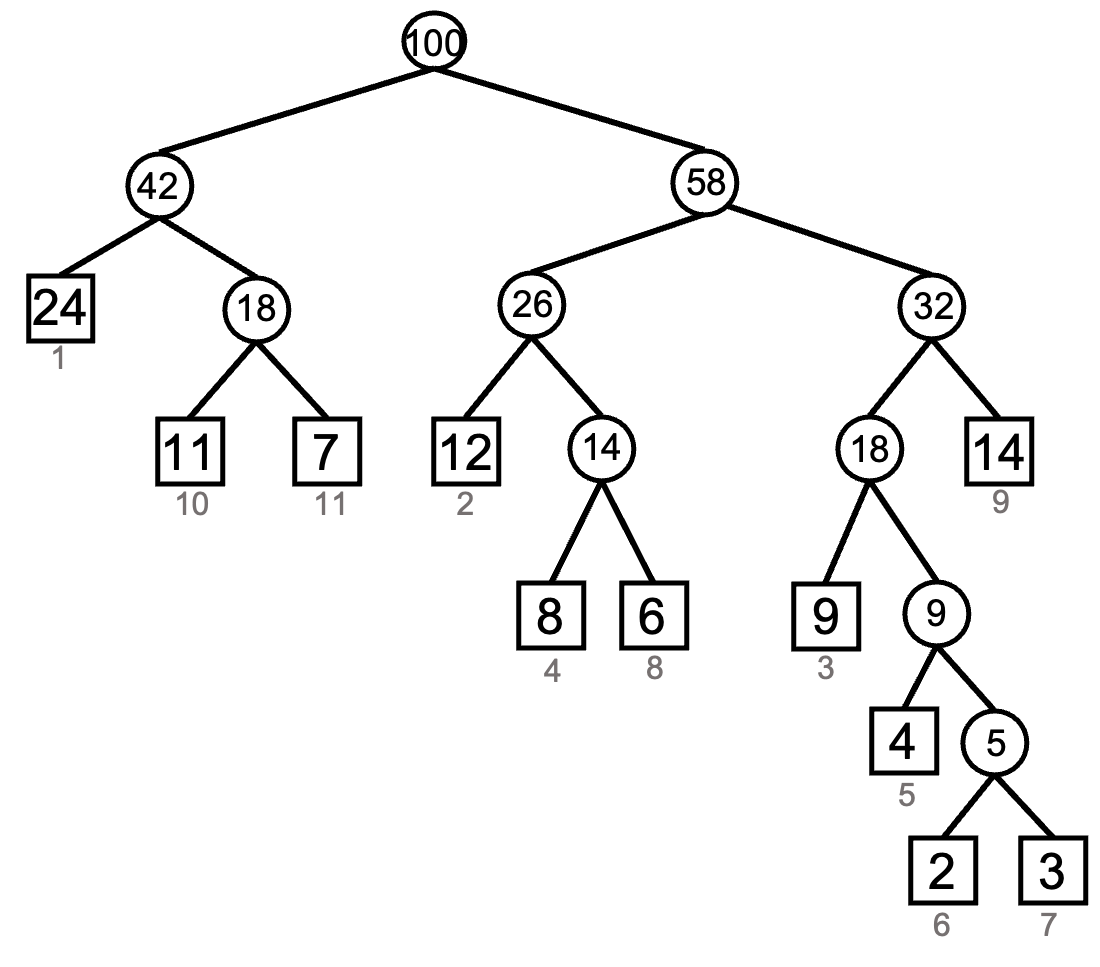} 
\caption{Intermediate tree built by the Hu-Tucker algorithm.}
\label{fig:tree:ht:intermediate}
\end{figure}

Now, all pairs of nodes are joinable, and thus the construction proceeds 
as in the construction of the Huffman tree, joining first \squared{24} and \nodeis{18},
then \nodeis{26} and \nodeis{32},
and finally \nodeis{42} and \nodeis{58}. The resulting tree is shown in Figure~\ref{fig:tree:ht:intermediate}.

Due to the joining of non-adjacent nodes, the alphabetic order of the leaves does not correspond anymore
to the original order. However, what is needed from this tree is only its structure, more precisely the lengths
$\mol{\ell_1,\ell_2,\ldots,\ell_n}$, where $\ell_i$ is the length of the root-to-leaf path for symbol $s_i$. 
It is possible to show that there exists a tree with leaves having such levels in that order, thus an alphabetic tree, 
whose cost is equal to the one built, and that such a cost is optimal. We refer
to \cite{hu1971optimal} for the details.

\begin{figure}[H]
\centering
\includegraphics[width=0.60\textwidth]{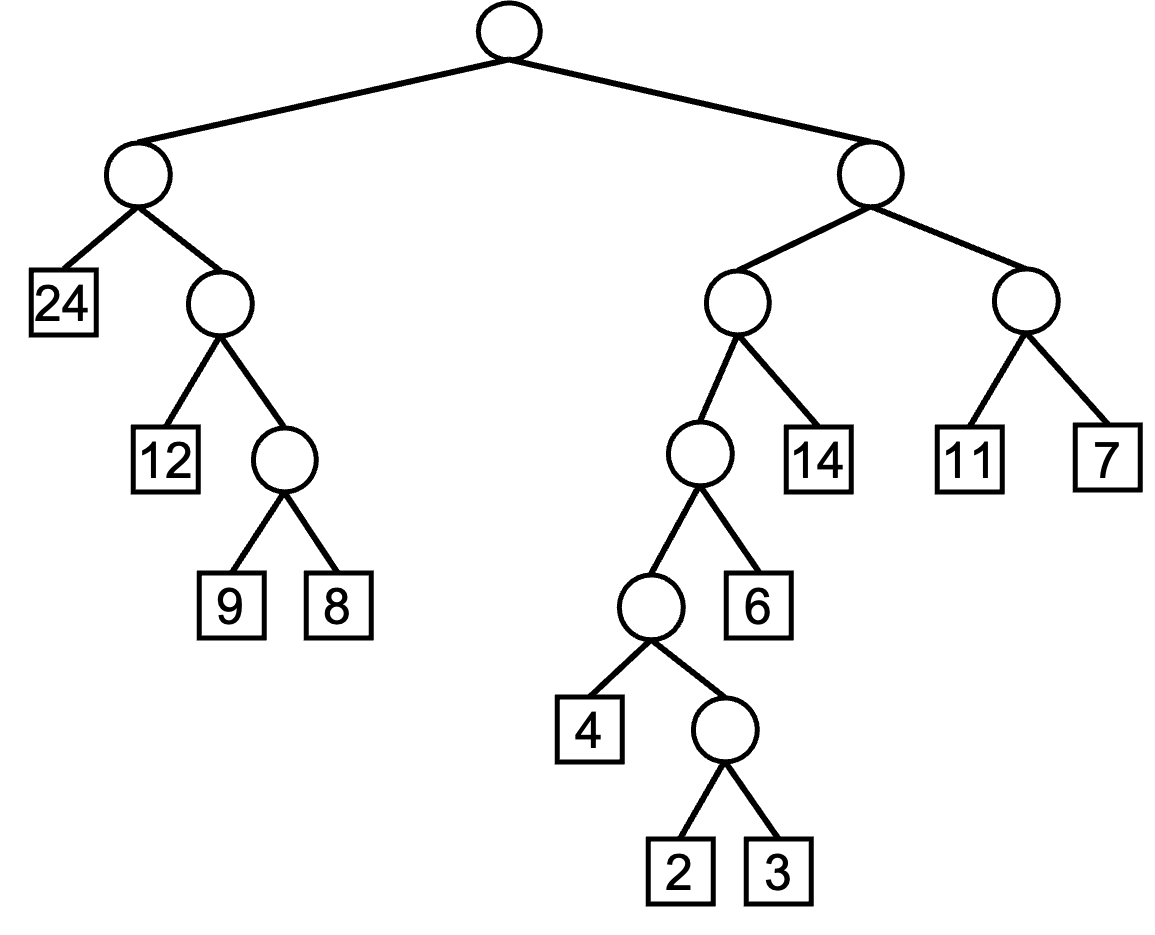} 
\caption{Final optimal alphabetic tree built by the Hu-Tucker algorithm.}
\label{fig:tree:ht:final}
\end{figure}

In our example, re-ordering the lengths to match the initial ordering of the symbols, we have $\mol{\ell_1,\ell_2,\ldots,\ell_{11}} = \mol{2,3,4,4,5,6,6,4,3,3,3}$
and the corresponding optimal alphabetic tree is shown in Figure~\ref{fig:tree:ht:final}. Notice that
knowing the lengths $\mol{\ell_1,\ell_2,\ldots,\ell_n}$ one can easily build the tree, just by using always
the leftmost available path.

For the reader's convenience, we summarise the\textit{ modus operandi} of the
Hu-Tucker algorithm in  the pseudocode Algorithm~\ref{alg:gw}.

\begin{algorithm}[H]\label{alg:gw}
\caption{Hu-Tucker algorithm}
\KwIn{Symbols $S=\{s_1,\dots,s_n\}$ and $P=\mol{p_1,\dots,p_n}$ the associated probability distribution.}

Build the intermediate tree $T'$ as follows. Start with an ordered forest of one-node trees corresponding to the $n$ probabilities. Consider two nodes of the list to be {\em joinable} if there are no internal nodes in between them.

\Repeat{there is only one tree} {
    Find the leftmost pair of joinable nodes that gives the smallest sum\;
    Merge the two nodes keeping the merged node in the position of the left node of the pair
}

Let $\ell_i$ be the the length of the root-to-leaf paths in $T'$ for symbol $s_i$

Build the final tree $T$ with leaves at levels $\mol{\ell_1,\ell_2,\ldots,\ell_n}$, 
with leaf $i$ associated to symbol $s_i$.

\KwOut{The tree $T$}
\end{algorithm}

We conclude this section by mentioning that a detailed implementation of 
the Hu-Tucker algorithm was given in \cite{Yoe,by}. Moreover, the 
procedure given in \cite{Yoe} finds a minimal cost tree whose longest
path length and total path length are minimal.

\subsection{Garsia and Wachs' algorithm}

The Garsia-Wachs algorithm is similar to the Hu-Tucker algorithm: it first builds an intermediate tree, and then uses the lengths of the
leaves in the intermediate tree to build the final alphabetic tree. The construction of the intermediate tree differs from the Hu-Tucker algorithm, although it is somewhat similar.
Perhaps the construction of the intermediate tree is somewhat simpler since there is no need to distinguish between joinable and non-joinable nodes.
The construction of the final tree from the intermediate tree is the same.
As in the Hu-Tucker algorithm, we start from the initial (ordered) list of probabilities and we perform $n-1$ steps in each of which we join two probabilities and
move the resulting node to an appropriate position in the new list. The rule used to select the two nodes to be joined is what differentiates the algorithm
from the Hu-Tucker construction.  The Garsia-Wachs algorithm joins the two rightmost consecutive nodes for which
the sum of the probabilities is the smallest. Then, the newly created element, which in the tree will be the parent of the two nodes that have been joined,
is moved to the right of the current position placing it just before the first node whose probability is greater or equal to its probability;
or at the end of the list if there is no such a node. More formally, let
$$\mol{p_1,p_2,\ldots\ldots\ldots,p_m}$$
the ordered list of probabilities for a generic step of the algorithm (where $m=n$ initially and will decrease by $1$ at each iteration).
Let $p_i,p_{i+1}$ be the rightmost consecutive probabilities whose sum is the minimum possible over all the 
consecutive pairs. Let $p_k$, $k\ge i+2$, be the first probability such that $p_k\ge p_i+p_{i+1}$. If such a probability exists, the new list of $m-1$ probabilities
is
$$\mol{p_1,\ldots,p_{i-1},p_{i+2},\ldots, p_{k-1},(p_i+p_{i+1}),p_k,p_{k+1},\ldots,p_m}.$$
If $p_k$ does not exist the new probability is moved to the end of the list, that is, the new list is
$$\mol{p_1,\ldots,p_{i-1},p_{i+2},\ldots, p_{m-1},p_m,(p_i+p_{i+1})}.$$
After $n-1$ steps, the intermediate tree is built.

\begin{figure}[h]
\centering
\includegraphics[width=0.6\textwidth]{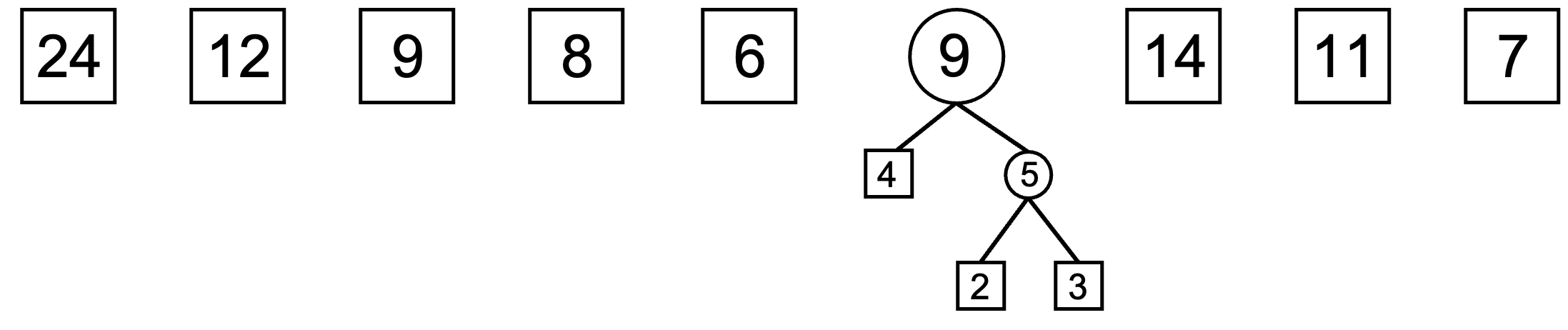} 
\caption{List of nodes after step 2 of the Garsia-Wachs algorithm}
\label{fig:tree:gw2}
\end{figure}

Let us clarify the construction with an example. We consider the same probability distribution $P$ that we have used for the previous examples.
The initial list is the same as for the Hu-Tucker algorithm, that is the one depicted in Figure~\ref{fig:initial:list}.
The consecutive pair of probabilities with the smallest sum is \squared{2} and \squared{3} and thus they get joined. Moreover, the first 
probability on the right side of the joined probability is \squared{6} thus the new node \nodeis{5} will be placed right before \squared{6},
leading to the same list of the Hu-Tucker algorithm depicted in Figure~\ref{fig:ht:step1}. For the next step, the Garsia-Wachs algorithm behaves
differently. Indeed the consecutive pair of probabilities whose sum is minimum is \squared{4} and \nodeis{5}. This creates the
new probability \nodeis{9} and the first probability greater than $9$ is \squared{14}. Thus, the new list is the one shown in Figure~\ref{fig:tree:gw2}.

\begin{figure}[h]
\centering
\includegraphics[width=0.6\textwidth]{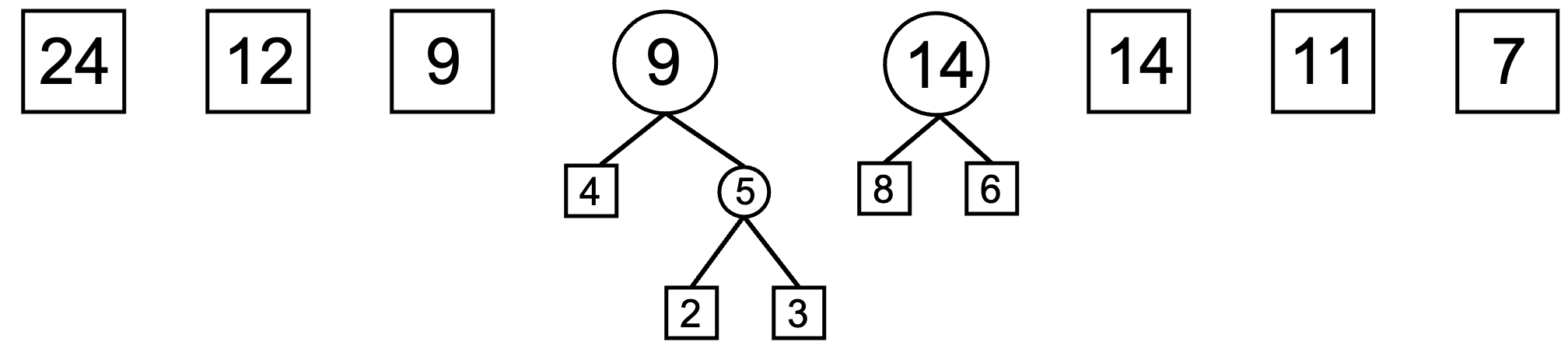} 
\caption{List of nodes after step 3 of the Garsia-Wachs algorithm}
\label{fig:tree:gw3}
\end{figure}

Now the smallest sum is given by \squared{8} and \squared{6} and the first probability greater or equal to their sum is \squared{14}, so the newly
created node \nodeis{14}, will be placed right before \squared{14}, as shown in Figure~\ref{fig:tree:gw3}.

\begin{figure}[h]
\centering
\includegraphics[width=0.55\textwidth]{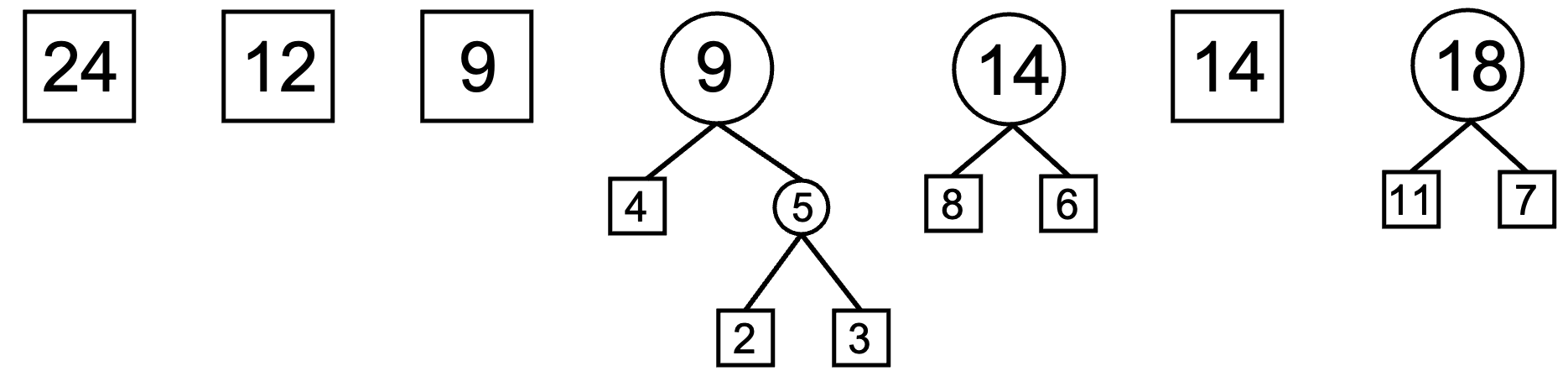} 
\caption{List of nodes after step 4 of the Garsia-Wachs algorithm}
\label{fig:tree:gw4}
\end{figure}

The rightmost smallest sum is now given by \squared{11} and \squared{7} and the new node will be placed at the end of the list, as shown in Figure~\ref{fig:tree:gw4} (Notice that also \nodeis{9} and \squared{9} give 18 as sum, but the algorithm
takes the rightmost pair).

\begin{figure}[h]
\centering
\includegraphics[width=0.45\textwidth]{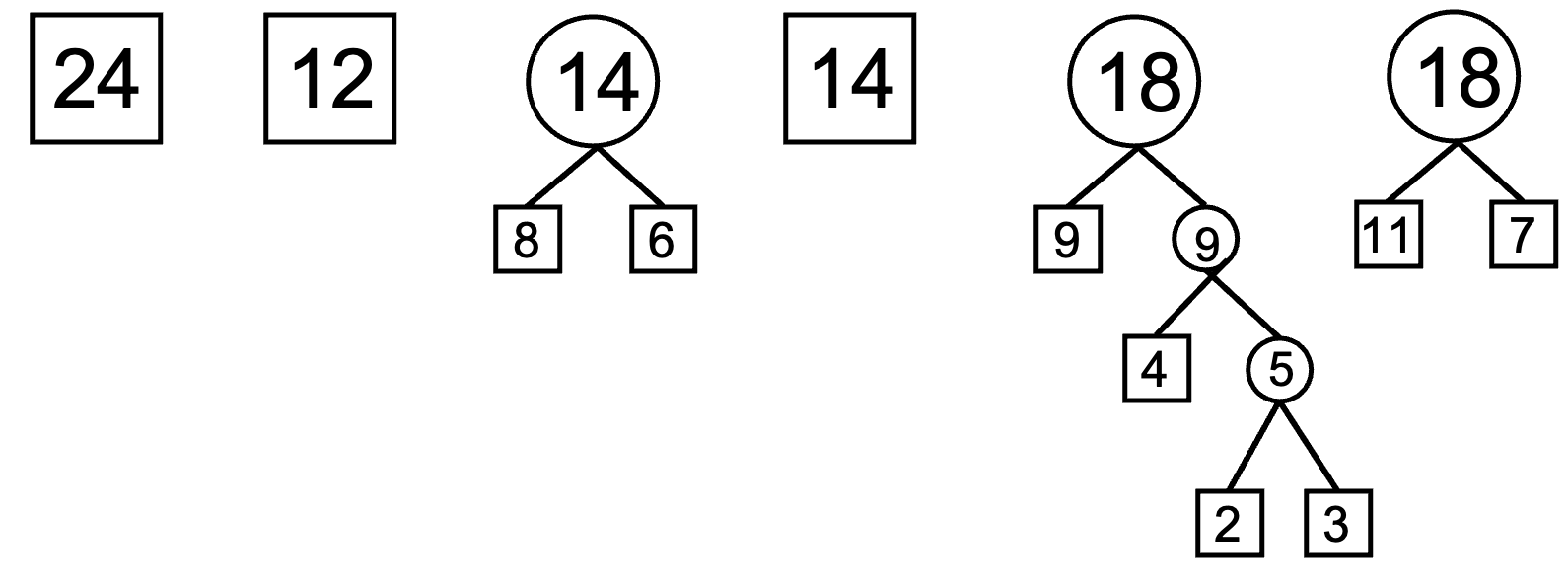} 
\caption{List of nodes after step 5 of the Garsia-Wachs algorithm}
\label{fig:tree:gw5}
\end{figure}

For the next step, the smallest sum is obtained by joining \squared{9} and \nodei{9}, an operation that creates the node \nodeis{18}. 
The first node to their right with a probability equal or greater than $18$ is the last node of the list \nodeis{18}, thus the newly created
node will be placed just before the last one, as depicted in Figure~\ref{fig:tree:gw5}.

\remove{
\begin{figure}
\centering
\begin{subfigure}{.5\textwidth}
  \centering
  \includegraphics[width=0.8\linewidth]{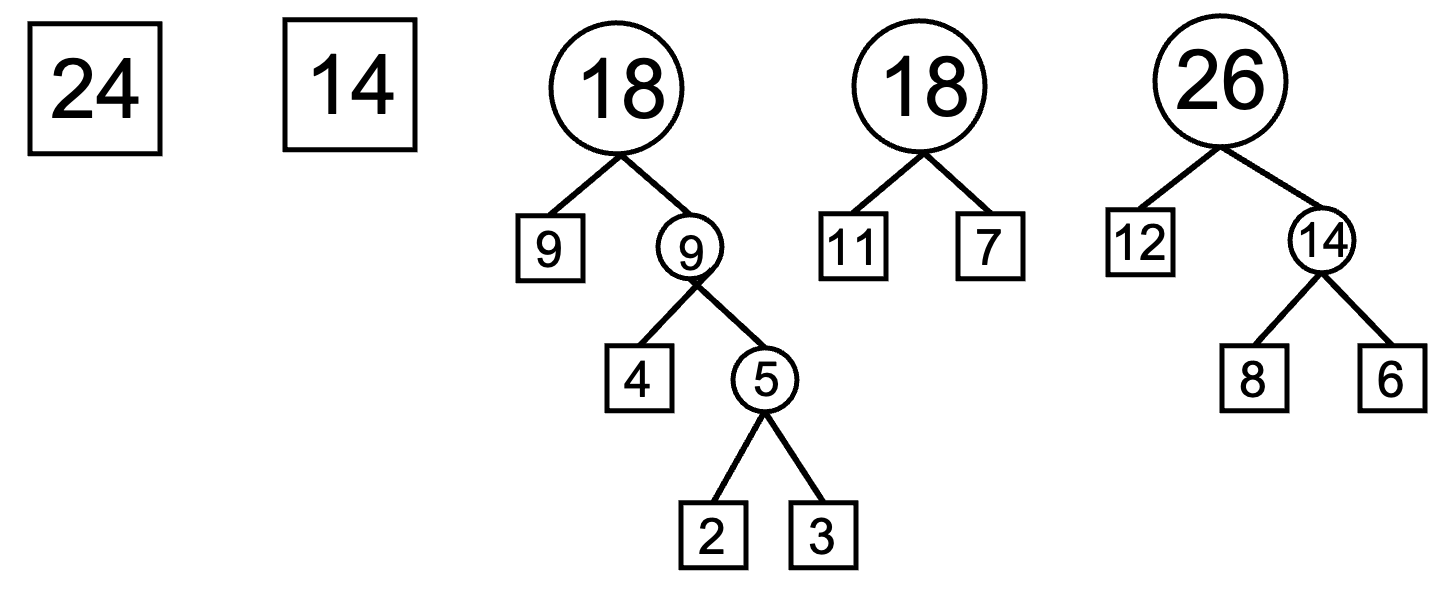}
  \caption{GW step 6}
  \label{fig:tree:gw6}
\end{subfigure}%
\begin{subfigure}{.5\textwidth}
  \centering
  \includegraphics[width=0.8\linewidth]{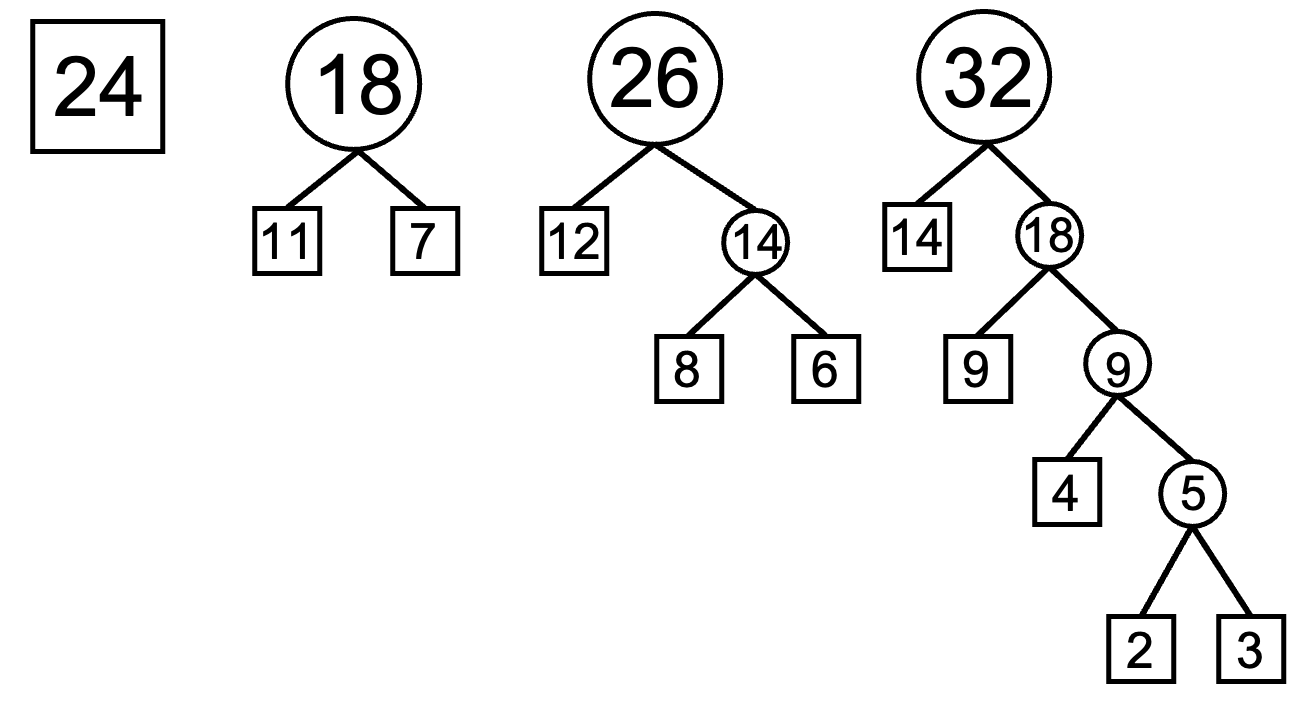}
  \caption{GW step 7}
  \label{fig:tree:gw7}
\end{subfigure}

\bigskip

\begin{subfigure}{.5\textwidth}
  \centering
  \includegraphics[width=0.8\linewidth]{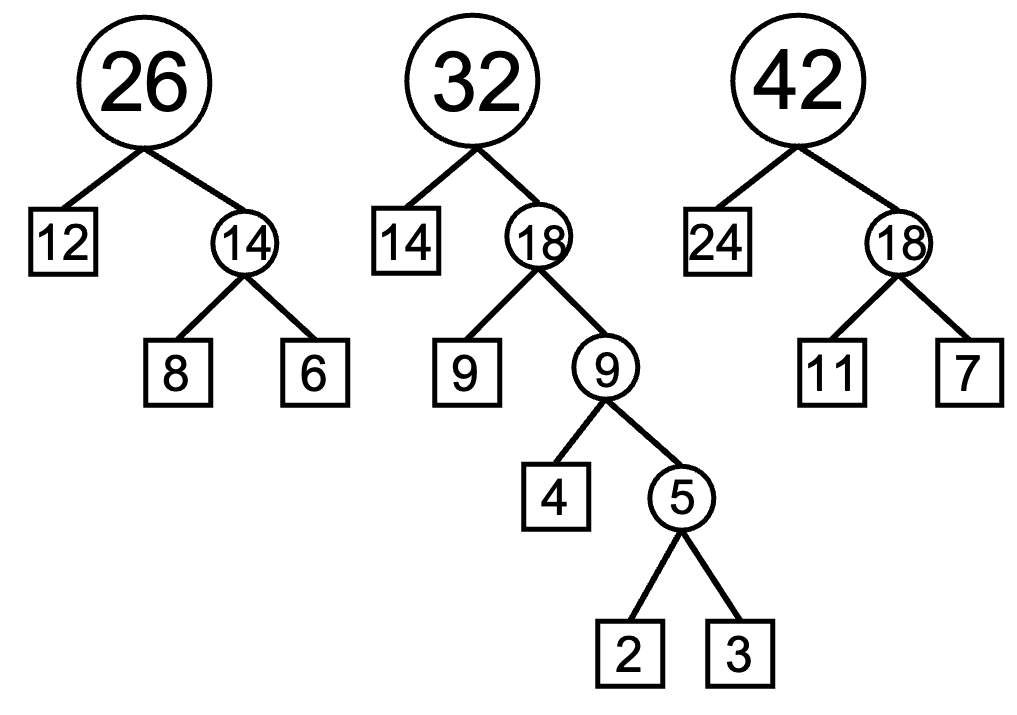}
  \caption{GW step 8}
  \label{fig:tree:gw8}
\end{subfigure}%
\begin{subfigure}{.5\textwidth}
  \centering
  \includegraphics[width=0.8\linewidth]{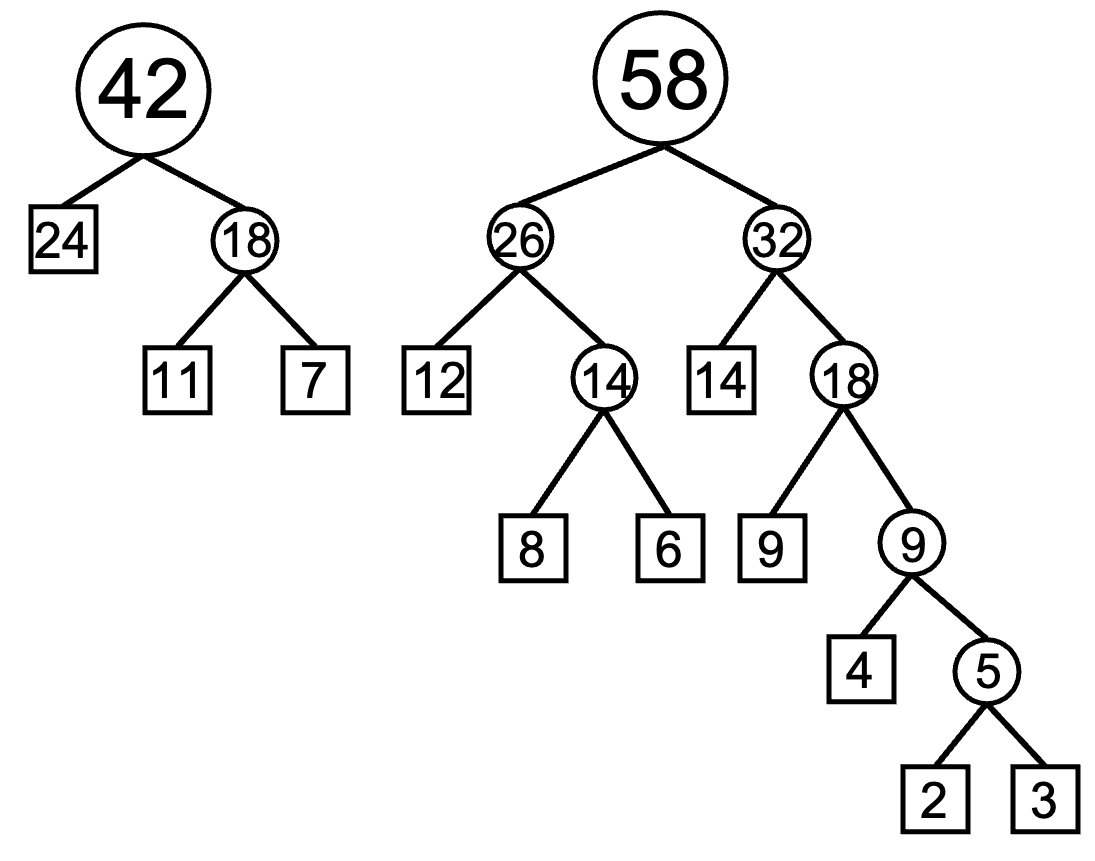}
  \caption{GW step 9}
  \label{fig:tree:gw9}
\end{subfigure}
\end{figure}
}

The next step will join \squared{12} and \nodeis{14} creating a new node \nodeis{26} that will be placed 
at the end of the list as shown in Figure~\ref{fig:tree:gw6}.

\begin{figure}[h]
\centering
\includegraphics[width=0.45\textwidth]{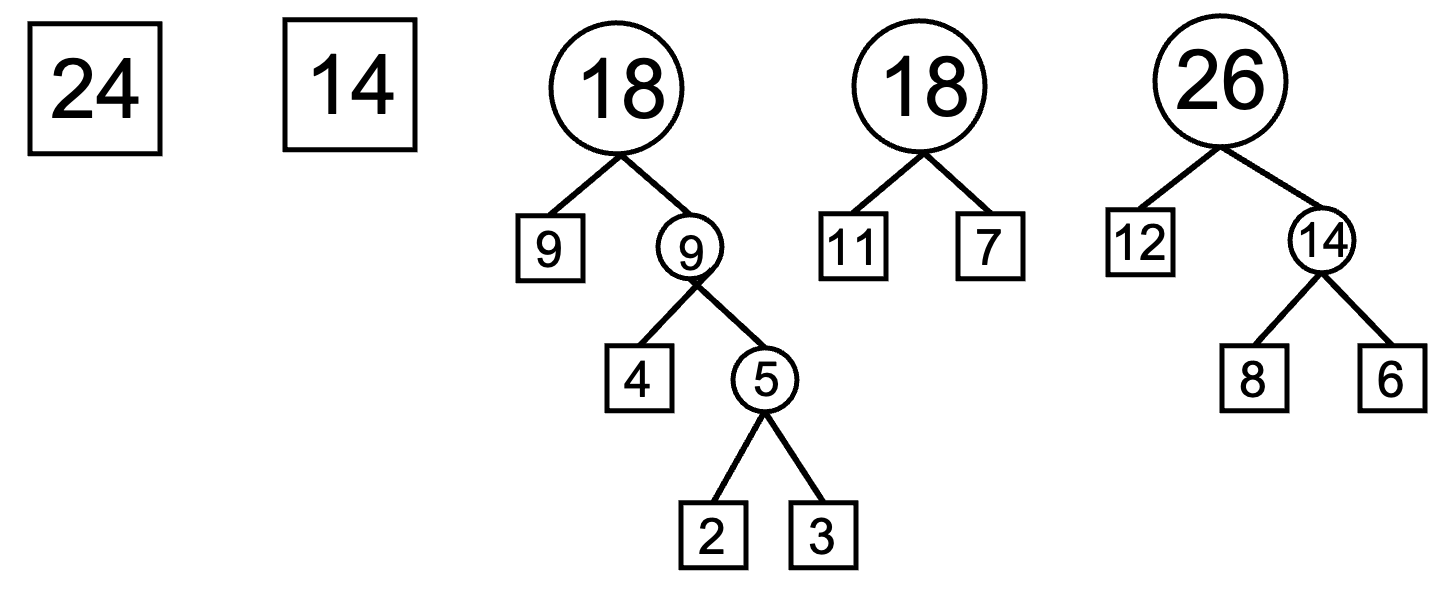} 
\caption{List of nodes after step 6 of the Garsia-Wachs algorithm}
\label{fig:tree:gw6}
\end{figure}

The next step will join \squared{14} and \nodeis{18} creating a new node \nodeis{32} that will be placed 
at the end of the list as shown in Figure~\ref{fig:tree:gw7}.

\begin{figure}[h]
\centering
\includegraphics[width=0.4\textwidth]{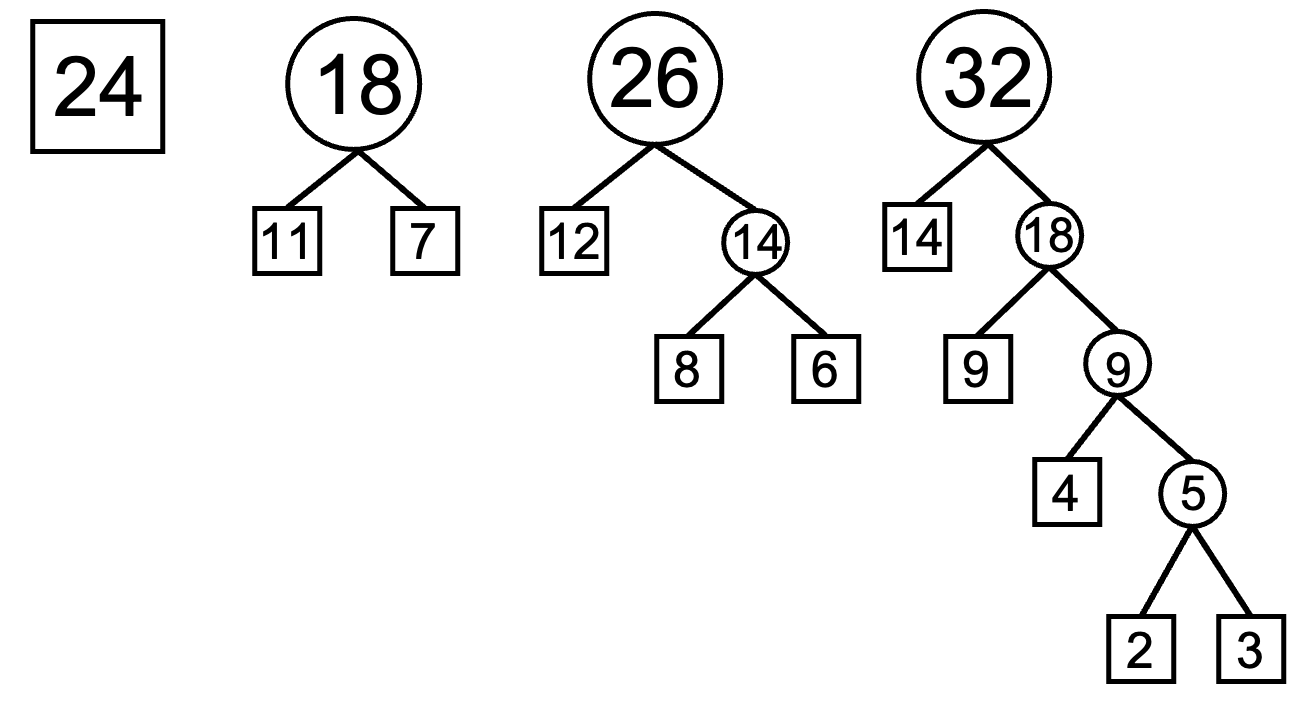} 
\caption{List of nodes after step 7 of the Garsia-Wachs algorithm}
\label{fig:tree:gw7}
\end{figure}

The subsequent step will join \squared{24} and \nodeis{18} creating a new node \nodeis{42} that will be placed 
at the end of the list as shown in Figure~\ref{fig:tree:gw8}.

\begin{figure}[h]
\centering
\includegraphics[width=0.35\textwidth]{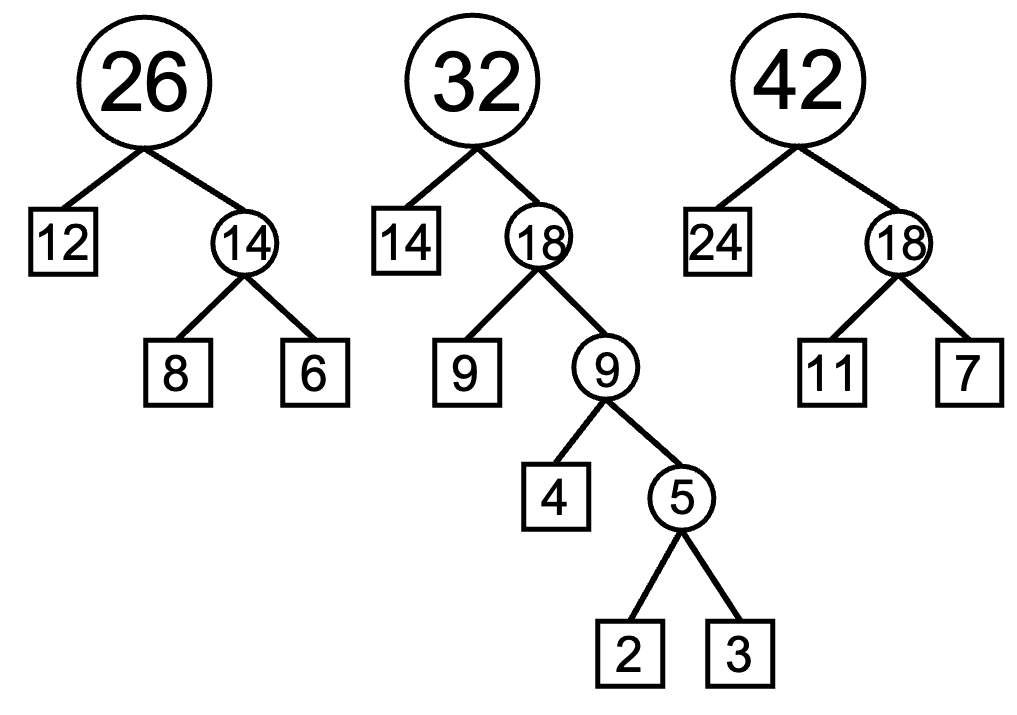} 
\caption{List of nodes after step 8 of the Garsia-Wachs algorithm}
\label{fig:tree:gw8}
\end{figure}

\begin{figure}[h]
\centering
\includegraphics[width=0.6\textwidth]{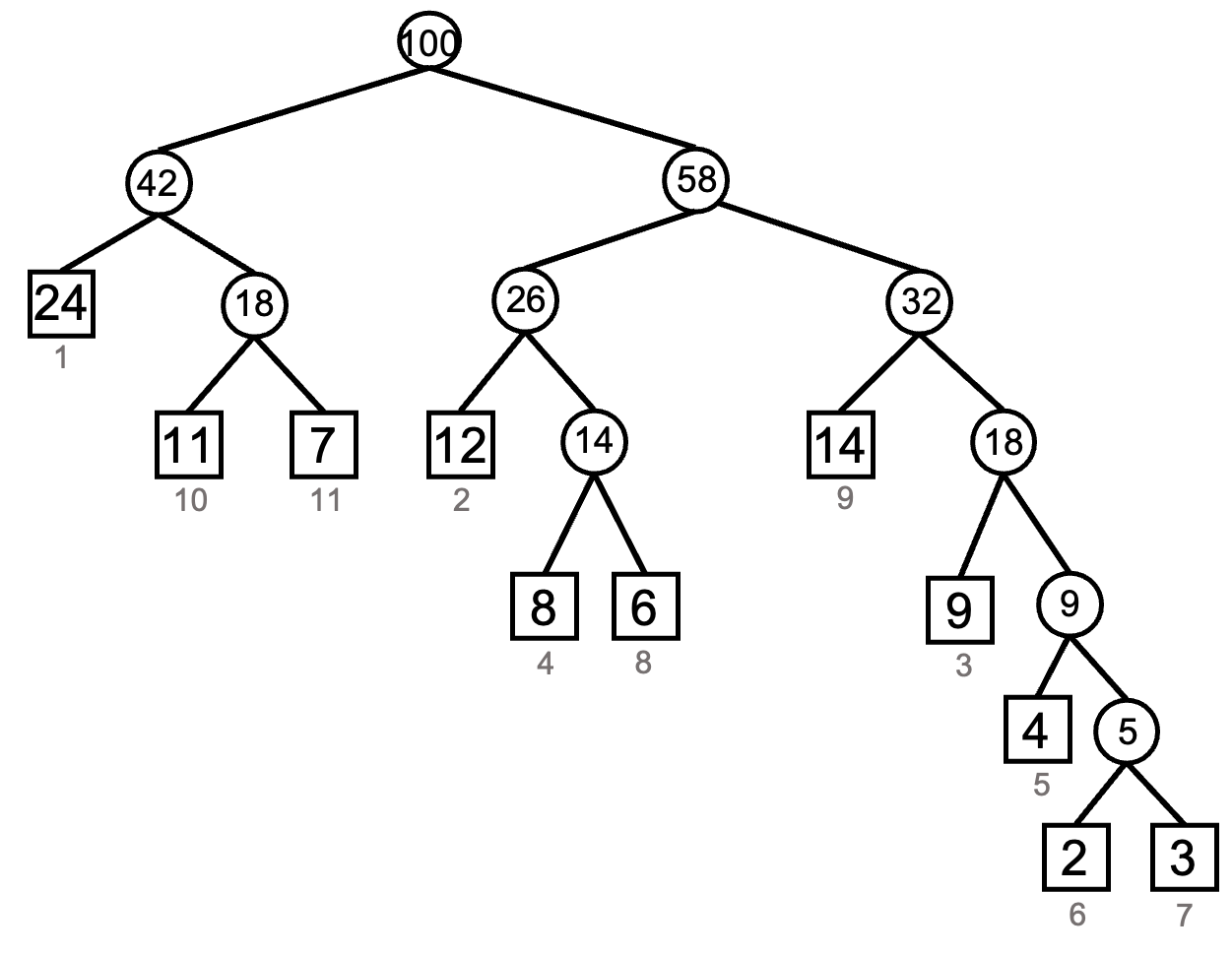} 
\caption{The intermediate tree  built by the Garsia-Wachs algorithm}
\label{fig:tree:gw:intermediate}
\end{figure}

Step 9 joins \nodeis{26} and \nodeis{32}, as shown in Figure~\ref{fig:tree:gw9}, and the final step gives the intermediate tree shown in~Figure~\ref{fig:tree:gw:intermediate}.

\begin{figure}[h]
\centering
\includegraphics[width=0.35\textwidth]{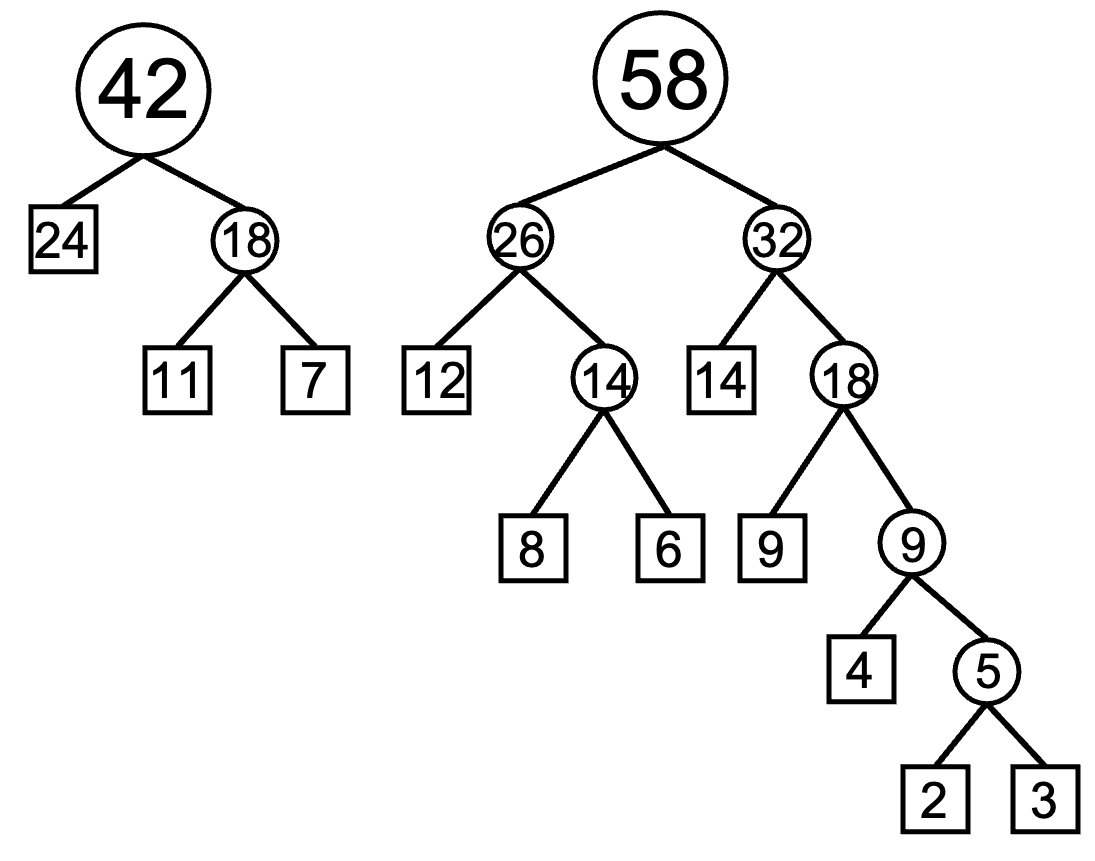} 
\caption{List of nodes after step 9 of the Garsia-Wachs algorithm}
\label{fig:tree:gw9}
\end{figure}

From this tree, as done by the Hu-Tucker algorithm, we extrapolate the lengths of the codewords associated with the symbols and we reorder them to match the initial ordering.
Node \squared{24} has length 2, node \squared{12} has length 3, node \squared{9} has length $4$, and so on, leading to the vector of lengths $\mol{\ell_1,\ldots\ell_{11}}=\mol{2,3,4,4,5,6,6,4,3,3,3}.$ This is the same length vector of the 
intermediate tree of the Hu-Tucker algorithm (although the intermediate trees are slightly different); hence the final
tree is the same as the one of the Hu-Tucker algorithm already shown in Figure~\ref{fig:tree:ht:final}.

As done for the Hu-Tucker algorithm, we summarise the\textit{ modus operandi} of the
Garsia-Wachs algorithm in the pseudocode Algorithm \ref{alg:hu-tucker}.

\begin{algorithm}[H]\label{alg:hu-tucker}
\caption{Garsia-Wachs algorithm}
\KwIn{Symbols $S=\{s_1,\dots,s_n\}$ and $P=\mol{p_1,\dots,p_n}$ the associated probability distribution.}
Build the intermediate tree $T'$ as follows.

Start with an ordered forest of trees with one node corresponding to the
$n$ probabilities.

\Repeat{there is only one tree} {
    Find the rightmost pair of consecutive nodes $p_i,p_{i+1}$ that gives the smallest sum
    in the current list $p_1,\ldots,p_m$\;

    Let $k\ge i+2$ be such that $p_k$ is the first probability satisfying $p_k\ge p_i+p_{i+1}$\;

    If such $k$ exists, the new list is $p_1,\ldots,p_{i-1},p_{i+2},\ldots, p_{k-1},(p_i+p_{i+1}),p_k,p_{k+1},\ldots,p_m.$

    If such $k$ does not exist, the new list is $p_1,\ldots,p_{i-1},p_{i+2},\ldots, p_{m-1},p_m,(p_i+p_{i+1}).$
}

Let $\ell_i$ be the the length of the root-to-leaf paths in $T'$ for symbol $s_i$

Build the final tree $T$ with leaves at levels $\mol{\ell_1,\ell_2,\ldots,\ell_n}$, 
with leaf $i$ associated with the symbol $s_i$.

\KwOut{The tree $T$}
\end{algorithm}

We conclude this section by mentioning that the paper \cite{Bird}
gave an $O(n\log n)$ implementation of the Garsia-Wachs algorithm
in the framework of functional programming.

 \section{Necessary and sufficient conditions for the existence of
 alphabetic codes}\label{sec:conditions}

Given a multiset of integers $\{\ell_1, \dots,\ell_n\}$, the well known Kraft 
inequality states that there exists a binary prefix code with codeword lengths 
$\ell_1, \dots,\ell_n$ if and only if it holds that
\begin{equation}\label{kraft}
\sum_{i=1}^n2^{-\ell_i}\leq 1.
\end{equation}
It is natural to ask whether similar conditions hold also for prefix and alphabetic codes.
As expected, the answer is positive, but the conditions are considerably
more complicated than (\ref{kraft}).

Since the ordering of the codeword-to-symbol association is a hard constraint on alphabetic
codes, we recall the general setup. Let $S=\{s_1, \ldots, s_m\}$  be a set of symbols and $\prec$ be a total order relation on $S$, that is, for which we have $s_1\prec \dots \prec s_m$.  Given a list of integers
$L = \mol{\ell_1, \dots,\ell_n}$, we ask under which conditions there exists
an alphabetic code $w:S\mapsto \{0,1\}^+$ that assigns a codeword of length
$\ell_i$ to the symbol $s_i$, for $i=1, \ldots, n$.

\remove{
 First, let us introduce a weaker condition shown by Ahlswede and Wegener in \cite{AW}. Given a list of positive integers $L = \mol{\ell_1, \dots,\ell_n}$, if 
 \begin{equation}
     \sum_{i=1}^n 2^{-\ell_i} \leq \frac{1}{2},
 \end{equation}
 then there exists an alphabetic code with codeword lengths $L$. However, this condition is sufficient but not necessary. Intuitively, the reason why such a condition is not necessary is that it does not take into account the ordering of lengths imposed by the alphabetic property.
}

The first necessary and sufficient condition for the existence of alphabetic codes 
was given by Yeung \cite{yeung1991alphabetic}. To properly describe Yeung's result,
we need to introduce some preliminary definitions.
Let $\cmap: (\mathbb{R_+}, \mathbb{R_+}) \to \mathbb{R_+}$ be a mapping defined as 
\begin{equation*}
    \cmap(a,b) = \left\lceil\frac{a}{b}\right\rceil b.
\end{equation*}

\begin{definition}[\cite{yeung1991alphabetic}]
    For any list of positive integers $L = \mol{ \ell_1, \dots,\ell_n }$, define the numbers $\yf(L, k)$ as
    \begin{equation*}
    \yf(L, k)= \begin{cases}
            0 & \text{ if } k = 0,\\
            \cmap(\yf(L, k-1), 2^{-\ell_k}) + 2^{-\ell_k} & \text{ if } 1\leq k \leq n.\\
        \end{cases}
    \end{equation*}
\end{definition}
 Yeung proved the following result.
\begin{theorem}[\cite{yeung1991alphabetic}]\label{th:yeung_condition}
    There exists a binary alphabetic code with codeword lengths $L = \mol{\ell_1, \dots,\ell_n}$ if and only if $\yf(L, n) \leq 1$.
\end{theorem}

A similar and equivalent condition was later provided by Nakatsu in \cite{nakatsu1991bounds}. We first recall the following definitions.

\begin{definition}
    For a binary fraction $x$ and integer $i\geq 1$,
    let the function $\trunc$ be defined as
    \begin{equation}\label{trunc}
    \trunc(i,x)=\frac{\lfloor 2^i\,x\rfloor}{2^i},
    \end{equation}
     that is, $\trunc(i,x)$ is the fraction obtained by considering only the first 
     $i$ bits in the binary representation of $x$.
\end{definition}

\begin{definition}[\cite{nakatsu1991bounds}]\label{def:nakatsu}
    Let $L = \mol{\ell_1,\dots,\ell_n}$ be a list of positive integers. 
    Let $\alpha_i = \min(\ell_{i-1},\ell_i)$, for $i=2,\dots,n$. 
    Define the following recursive  function $\sm$ as
    \begin{equation}\label{sm}
        \sm(L, i) = \begin{cases}
            \trunc(\alpha_i, \sm(L, i-1))+2^{-\alpha_i} & \text{if } i \geq 2, \\
            0 & \text{if } i = 1.
        \end{cases}
    \end{equation}
\end{definition}
Nakatsu proved the following result.
\begin{theorem}[\cite{nakatsu1991bounds}]\label{th:nakatsu_condition}
    There exists a binary alphabetic code with codeword lengths $L = \mol{\ell_1, \dots,\ell_n}$ if and only if $\sm(L, n) < 1$.
\end{theorem}

Subsequently, a different necessary and sufficient condition was introduced by Sheinwald in \cite{sheinwald1992binary}. As for the previous conditions, we need to introduce some preliminary definitions.
\begin{definition}[\cite{sheinwald1992binary}]
    Let $L = \mol{\ell_1,\dots,\ell_n}$ be a list of positive integers. For a binary fraction $x$ and integer $i\geq 1$,
    let the function $\tf$ be defined as
    \begin{equation*}
        \tf(i,x) = \trunc(i,x) + \frac{\lceil x - \trunc(i,x) \rceil}{2^i},
    \end{equation*}
    that is, $\tf(i,x)$ is equal to $x$ if $\trunc(i,x)=x$, and to $\trunc(i,x) + 2^{-i}$ otherwise. Moreover, let $\varphi$ be the following function
    \begin{equation*}
        \varphi(L, i) = \begin{cases}
             \tf(\ell_i, \varphi(L,i-1)) +  2^{-\ell_i} & \text{ if } 2 \leq i \leq n, \\
            2^{-\ell_1}                                 & \text{ if } i = 1.
        \end{cases}
    \end{equation*}
\end{definition}
Sheinwald provided the following result.
\begin{theorem}[\cite{sheinwald1992binary}]\label{th:sh}
    There exists a binary alphabetic code with codeword lengths $L = \mol{\ell_1, \dots,\ell_n}$ if and only if $\varphi(L, n) \leq 1$.
\end{theorem}
Although clearly equivalent, there might be scenarios where one of the conditions stated in Theorems \ref{th:yeung_condition}, \ref{th:nakatsu_condition}, 
and \ref{th:sh} could be more easily applicable than the others.

\medskip
We conclude this section with an extension of the above results
to the case where the codewords of the alphabetic code must respect some additional constraints \cite{depriscopersiano}. 
Namely, by looking at the lengths of the codewords $\ell_i$ as the lengths of the root-to-left paths in the
tree that represents the code, one can consider the number of ``left'' edges, $l_i$ and
the number of ``right'' edges $r_i$, whose sum gives $\ell_i=l_i+r_i$. We can define
the path vector $\bar{v}$ as the ordered set of pairs of $\mol{(l_1,r_1),(l_2,r_2),...,(l_n,r_n)}$. The question is:  Determine whether or not an alphabetic
code with a given path vector $\bar{v}=\mol{(l_1,r_1),(l_2,r_2),...,(l_n,r_n)}$
exists.

As for the previous condition, also in this case we need to introduce a specific notation.
Let $N=\max \{l_1+r_1,l_2+r_2,...,l_n+r_n\}$ and let
$F$ be a full tree of order $N$. Number the leaves of $F$ from $0$ through $2^N-1$. 
Define the {\em projection} of a node $u$ of $F$ as the set of leaves descendent of $u$. 
A projection is identified by the pair of indexes corresponding
to the leftmost and the rightmost leaf of the projection.

Consider the set of binary strings belonging to $\{0,1\}^{l+r}$, that is,
the set of strings consisting of $l$ bits equal to zero and $r$ bits equal to one; denote such a set by $\calS(l,r)$. 

Each element of $\gamma\in\calS(l,r)$ represents an $(l,r)$-node of $F$: the node whose path,
encoded with a $0$ for a left edge and with a $1$ for a right edge, gives $\gamma$.

Let $u$ be the node of $F$ identified by some element $\alpha$ of
$\calS(l_u,r_u)$. The projection of $u$ is given by $(a,b)$ where $a$ and $b$ are the integers whose binary representations
(with possible leading zeros) are respectively $\gamma\!\!\!\underbrace{00...00}_{N-(l_u+r_u)}$ and
$\gamma\!\!\!\underbrace{11...11}_{N-(l_u+r_u)}$.

A natural way to construct a binary tree with a given path vector is the following: for $k=1,2,...,n$, choose the leftmost available $(l_k,r_k)$-node of $F$ to be the $k^{th}$ leaf of the binary tree.

This strategy can be formalized as follows. Let $\B_0=0$ and for each $i=1,2,...,n$,
let $\gamma_i$ be the smallest element of $\calS(l_i,r_i)\cup \{\infty\}$ such that $\gamma_i 2^{n-(l_i+r_i)} > \B_{i-1}$.
Then define $\A_i=\gamma_i 2^{N-(l_i+r_i)}$ and $\B_i=2^{N-(l_i+r_i)}(\gamma_i+1)-1$.
Notice that the binary representation of $\A_i$ is $\gamma_i\!\!\!\underbrace{00...00}_{N-(l_i+r_i)}$
and the one of $\B_i$ is $\gamma_i\!\!\!\underbrace{11...11}_{N-(l_i+r_i)}$ and
thus $(\A_i,\B_i)$ is the projection of a $(l_i,r_i)$-node (provided that a tree with path vector $\bar{v}$ exists).
From the
definition it follows that $\B_k < \A_{k+1}$,
that is, the projections are disjoint and in increasing order.
In paper \cite{depriscopersiano} the authors proved the following result.

\begin{theorem}[\cite{depriscopersiano}]\label{th:depriscopersiano}
Let $\bar{v} =\mol{(l_1,r_1),(l_2,r_2),...,(l_n,r_n)}$ be a vector of pairs
of positive integers.  A binary tree with path vector $\bar{v}$ exists
if and only if
\begin{equation*}
\B_n < 2^N.
\end{equation*}
\end{theorem}
We notice that if $\B_n$ cannot be defined then the condition of the theorem is
not satisfied. Conversely, if $\B_n$ can be defined then it is surely strictly less
than $2^N$, thus a binary tree with path vector $\bar{v}$ exists
if and only if $\B_n$ can be defined.

\section{Upper bounds on the average length of optimal alphabetic codes}\label{sec:upp}
For practical and theoretical reasons, it is often important to know an estimate
of the minimum average length of alphabetic codes \textit{before} building them, that is,
in terms of a closed formula of the symbol probabilities alone.
In this section, we review the relevant literature on the topic. 

We first recall that,  as discussed in Section \ref{sec:optimal},  
optimal alphabetic codes can be constructed in 
time $O(n\log n)$. 
Most of the studies that provide
 upper bounds on the average length of optimal alphabetic codes do so by 
the following procedure:
\begin{itemize}
    \item first,  they design
linear-time construction algorithms for sub-optimal codes, 
\item subsequently, they compute explicit  upper bounds on the constructed codes, in terms
of some partial information on the probability distribution of the set of symbols.
\end{itemize}
Clearly,   the derived bounds constitute  upper
bounds on the length of \textit{optimal} alphabetic codes, as well.

The first result was obtained by Gilbert and Moore \cite{gilbert1959variable} 
who proposed a linear time algorithm 
to construct an alphabetic code for a set 
of symbols $S$, with associated probability distribution $P$, whose average length is less than $H(P)+2$. Here, $H(P)=-\sum_{i=1}^n p_i\log p_i$ is the Shannon entropy of the distribution $P$. Let us briefly recall the idea of the algorithm:

\bigskip
\begin{algorithm}[H]\label{alg:gilbert}
\caption{Gilbert and Moore's algorithm}
Let $S=\{s_1,\dots,s_n\}$ be a set of symbols and $P=\mol{p_1,\dots,p_n}$ the associated probability distribution.

Compute 
\begin{equation*}
    r_i = \sum_{j=1}^{i-1} p_j + \frac{p_i}{2},\quad \forall i=1,\dots,n.
\end{equation*}

For each $i=1,\dots,n$, take the first $\lceil -\log p_i\rceil + 1$ bit of the binary expansion of $r_i$ to construct the codeword of the symbol $s_i$.
\end{algorithm}

\bigskip
The algorithm is straightforward. Intuitively, its correctness is due to the increasing value of the $r_i$'s, for $i=1, \ldots, n$,  that ensures the prefix and alphabetic properties of the constructed codewords. Moreover, since the codeword lengths are explicitly given, one can see that they satisfy the conditions presented in Section \ref{sec:conditions} for the existence of an alphabetic code with such lengths. Formally, we can summarize the result as follows.
\begin{theorem}[\cite{gilbert1959variable}]\label{th:gilbert_moore}
    For any set of symbols $S=\{s_1,\dots,s_n\}$, $s_1 \prec \dots \prec s_n$, with associated probabilities $P=\mol{p_1,\dots,p_n}$, Algorithm
    \ref{alg:gilbert} constructs an alphabetic code $C$ for $S$
    whose average length 
    $\mathbb{E}[C]$ satisfies
    \begin{equation}\label{GM}
        \mathbb{E}[C] = \sum_{i=1}^n p_i \ell_i < H(P) + 2,
    \end{equation}
    where $\ell_i$ is the length of the $i^{th}$ codeword.  Algorithm
    \ref{alg:gilbert} runs in linear time.
\end{theorem}
Note that since the average length of any prefix   code is lower bounded by $H(P)$, 
(and, therefore, \textit{a fortiori}, the average length of any 
alphabetic code is also lower bounded by $H(P)$)
one gets that Gilbert and Moore's codes are at most two bits away from the optimum. 

It is interesting to remark that the upper bound of Theorem \ref{th:gilbert_moore} 
cannot be improved unless one has some additional information on the probability distribution $P$ of the symbols.
In fact, one can see that 
the average length of the best alphabetical code 
for a set of three symbols $s_1\prec s_2\prec s_3$, with the probability distribution 
$P=\mol{\epsilon, 1-2\epsilon, \epsilon}$, is equal to $2-\epsilon$. On the other hand, the entropy $H(P)$ of the distribution $P$ can be arbitrarily small, as $\epsilon\to 0$.

Successively, Horibe \cite{HORIBE1977148} provided a better upper bound than the  $H(P)+2$  bound
of Gilbert and Moore, by giving an algorithm to construct 
alphabetic codes of average length less than 
\begin{equation}\label{ho}
    H(P)+2-(n+2)p_{\min}, 
\end{equation}
where $p_{\min}$ is the smallest probability of $P$. 
A naive implementation of the algorithm given in \cite{HORIBE1977148} has a quadratic time complexity and  Walker and Gottlieb \cite{WG} designed a more efficient
$O(n\log n)$ time algorithm. Successively, 
Fredman \cite{fred} gave a clever method that reduced the
time complexity to $O(n).$

However, Horibe's algorithm unlike Algorithm \ref{alg:gilbert}, does not explicitly specifies the codeword lengths. Indeed, it is  a \textit{weight-balancing} algorithm, 
similar to the classical Fano algorithm 
for sub-optimal prefix codes \cite[p.17]{sha}.
Horibe's algorithm constructs a binary tree whose root-to-leaf paths represent
the codewords of the alphabetic tree (as illustrated in Section \ref{AlpSear}).
The root of the tree is associated with the whole probability distribution
$P=\mol{p_1,\dots, p_n}$. Successively,  
one computes the  index $k$ that partitions the probabilities into the two sequences
$\mol{p_1, \ldots , p_k}$ and $\mol{p_{k+1}, \ldots, p_n}$, where $k$ is chosen in such a 
way that
\begin{equation}\label{k}
    \left |\sum_{z=1}^k p_z - \sum_{z=k+1}^n p_z\right | = \min_{1\leq\ell<n} \left |\sum_{z=1}^{\ell} p_z - \sum_{z=\ell+1}^n p_z\right |.
\end{equation}
The sequence $\mol{p_1, \ldots , p_k}$ is associated to the left child of the root,
and the sequence $\mol{p_{k+1}, \ldots, p_n}$ is associated to the right child of the root.

The process is successively iterated in $\mol{p_1, \ldots , p_k}$ and $\mol{p_{k+1}, \ldots, p_n}$.
In general, 
for any consecutive sequence of probabilities $\mol{p_i,\dots,p_j}$, associated to
a node $x$ in the tree, one computes
 the index  $k_{ij}$  such that 
\begin{equation*}
    \left |\sum_{z=i}^{k_{ij}} p_{z} - \sum_{z=k_{ij}+1}^j p_{z}\right | = \min_{i\leq \ell<j} \left |\sum_{z=i}^{\ell} p_{z} - \sum_{z=\ell+1}^j p_{z}\right |.
\end{equation*}
Successively, the sequence $\mol{p_i, \ldots , p_{k_{ij}}}$ is associated to the left child of the 
node $x$,
and $\mol{p_{k_{ij}+1}, \ldots, p_j}$ is associated to the right child of the node $x$.
The process is iterated till one gets sequences made by just \textit{one} element.

One can see that the binary tree constructed by such an algorithm is a valid alphabetic code. Furthermore, Horibe proved the following result.

\begin{theorem}[\cite{HORIBE1977148}]
    For any set of symbols $S=\{s_1,\dots,s_n\}$, $s_1 \prec \dots \prec s_n$, with associated probabilities 
    $P=\mol{p_1,\dots,p_n}$, the alphabetic code $C$ for $S$ constructed by Horibe's  weight balancing algorithm  has an average length $ \mathbb{E}[C]$
    upper bounded by 
    \begin{align*}
        \mathbb{E}[C] &\leq H(P) + 
        \sum_{i=1}^{n-1} \max(p_i,p_{i+1}) - p_{\min}\\
        & \leq H(P) +2-(n+2)p_{\min},
    \end{align*}
    where $p_{\min} = \min_{i} p_i$.
\end{theorem}

Yeung \cite{yeung1991alphabetic} improved the upper bound of Gilbert and Moore
(\ref{GM}), by designing an algorithm 
that produces an alphabetic code whose   average length is upper bounded by 
$$
H(P) + 2 - p_1 - p_n,
$$ 
where $p_1$ and $p_n$ are the probabilities of the first and the last symbol of the ordered set of symbols $S$, respectively. For such a purpose, Yeung proved that given a probability distribution $P=\mol{p_1,\dots,p_n}$, the codeword lengths $L=\mol{\ell_1,\dots,\ell_n}$ defined as 
\begin{equation}
        \ell_i = \begin{cases}
            \lceil-\log p_i \rceil & \text{ if } i=1 \text{ or } i=n,\\
            \lceil-\log p_i \rceil +1 & \text{ otherwise, }
        \end{cases}
\end{equation}
satisfy the condition of Theorem \ref{th:yeung_condition}. Therefore, 
one knows that there exists an alphabetic code with lengths $L=\mol{\ell_1,\dots,\ell_n}$
(equivalently, that there exists a binary tree whose leaves, read from left to right,
appear at the levels $\ell_1, \ldots , \ell_n,$ respectively).  Moreover, 
Yeung also provided a linear time algorithm to construct such a code. Let us briefly recall it.

\bigskip
\begin{algorithm}[H]\label{alg:yeung}
\caption{Yeung's algorithm}
Let $S=\{s_1,\dots,s_n\}$ be a set of symbols and $P=\mol{p_1,\dots,p_n}$ the associated probability distribution.

Compute the lengths $L=\mol{\ell_1,\dots,\ell_n}$ as follows
\begin{equation*}
        \ell_i = \begin{cases}
            \lceil-\log p_i \rceil & \text{ if } i=1 \text{ or } i=n,\\
            \lceil-\log p_i \rceil +1 & \text{ otherwise, }
        \end{cases}
\end{equation*}

For each $i=1,\dots,n$, choose the leftmost available 
leaf at the level  $\ell_i$ to be the codeword of the symbol $s_i$.
    
\end{algorithm}

\bigskip
Since the lengths in the above algorithm are explicitly defined,
one can quite easily get the following result.

\begin{theorem}[\cite{yeung1991alphabetic}]
    For any set of symbols $S=\{s_1,\dots,s_n\}$, $s_1 \prec \dots \prec s_n$, with associated probabilities $P=\mol{p_1,\dots,p_n}$, there exists an alphabetic code $C$ for $S$, that can be constructed in linear time and whose average length satisfies
    \begin{align*}
        \mathbb{E}[C] &\leq H(P) + 2 - p_1 \left (2-\log p_1 -\lceil -\log p_1 \rceil\right)
        - p_n \left (2-\log p_n -\lceil -\log p_n \rceil\right)\\
        &\leq H(P) + 2 - p_1 - p_n.
    \end{align*}
\end{theorem}

In \cite{yeung1991alphabetic} Yeung also proved the following relation between alphabetic codes and Huffman codes \cite{huffman1952method}.

\begin{theorem}[\cite{yeung1991alphabetic}]
    For any set of symbols $S=\{s_1,\dots,s_n\}$, $s_1 \prec \dots \prec s_n$, with associated probabilities $P=\mol{p_1,\dots,p_n}$, if the probabilities $P$ are in ascending or descending order, then the average length 
    of an optimal alphabetic code for $S$ is equal to the average length 
    of the Huffman code for $S$.
    \remove{
    , that is, 
    \begin{equation*}
        L_{\min}= L_{{\tt Huff}}.
    \end{equation*}
    }
\end{theorem}

One of the consequences of Theorem \cite{yeung1991alphabetic} is that one can use
the known upper bounds on the average length of Huffman codes
(e.g., \cite{julia}) to obtain upper bounds
on the average lengths of optimal alphabetic codes, \textit{when} the probability
distribution $P$ is ordered. In general, these upper bounds
are much tighter than the known upper bounds for alphabetic codes that hold for 
arbitrary probability distributions (i.e., not necessarily ordered).
Bounds on the length of Huffman codes as functions of partial
knowledge of the probability distribution have been widely studied, e.g., ~\cite{julia,BHSM19,CD89huffman,CD91huffman,CGT86,DD96huffman,DD97huffman,Gallager78,Johnsen80,MA87,YY02,Yeung91huffman}.

Following the line of work that concerns upper bounds
on the average lengths of optimal alphabetic codes,  
we mention the paper  \cite{nakatsu1991bounds} by Nakatsu, who claimed another 
upper bound on the minimum average length of alphabetical codes. 
Nakatsu method's requires the construction (as a preliminary step) 
of a Huffman code for the probability
distribution $P$, successively one suitably modifies the Huffman code in order
to obtain an alphabetic code for the same probability distribution $P$. Unfortunately, Sheinwald  \cite{sheinwald1992binary} pointed out a gap in the analysis carried out in \cite{nakatsu1991bounds}. It is not clear whether 
Nakatsu analysis' can be repaired. On the other hand, Fariña \textit{et al.} \cite{Farina} provided a multiplicative factor approximation. Indeed, they designed an algorithm to build an alphabetic code whose average length is at most
a factor of $1+O(1/\sqrt{\log |S|})$ more than the optimal one, where $|S|$ is the cardinality of the set of symbols $S$.

De Prisco and De Santis  \cite{de1993binary} gave a  new upper bound on the minimum average length of alphabetical codes. However, Dagan \textit{et al.} \cite{dagan} 
pointed out an issue  in the analysis in 
\cite{de1993binary} and proposed a 
modified upper bound bound. The idea in \cite{dagan} can be summarized as follows: 
\begin{itemize}
    \item From the initial probability distribution $P=\mol{p_1,\dots,p_n}$, compute the  \textit{extended} distribution $Q=\mol{0,p_1,0,\dots,0,p_n,0}$;
    \item apply the classic algorithm of Gilbert and Moore \cite{gilbert1959variable} to construct an alphabetic code $C$ for the distribution $Q$, 
    whose average length is less than $H(Q)+2=H(P)+2$; 
    \item prune the binary tree representing the alphabetical code $C$ by eliminating
    the leaves associated with  the null probabilities in $Q$, re-adjust the obtained 
    tree.
\end{itemize}
Dagan \textit{et al.}  proved  that the average length $\mathbb{E}[C]$ of the 
obtained alphabetic code  satisfies the following inequality:

\begin{align}\label{eq:dagan_bound}
    \mathbb{E}[C] &\leq H(P) + 2 -p_1 - p_n - \sum_{i=1}^{n-1} \min(p_i,p_{i+1})\nonumber\\
    &= H(P)+1-\frac{p_1+p_n}{2}+\frac{1}{2}\sum_{i=1}^{n-1}|p_i-p_{i+1}|.
\end{align}
Recently, the same approach  has been analyzed by Bruno \textit{et al.} \cite{noi}, who first pointed out an issue in the analysis in \cite{dagan} and
subsequently improved the upper bound (\ref{eq:dagan_bound}).
More precisely, Bruno \textit{et al.} \cite{noi} designed a linear time algorithm
for constructing an alphabetic code whose
average is upper bounded by 
a quantity smaller than (\ref{eq:dagan_bound}). Let us briefly recall the idea of the algorithm proposed in \cite{noi},
summarized in Algorithm~\ref{alg:ours}. {For such a purpose, we need to recall the following intermediate result.}
{
\begin{lemma}[\cite{noi}]\label{lemma:linear_time_alg}
    Let $L=\langle\ell_1,\dots,\ell_n\rangle$ be a list of integers, associated with the ordered symbols $s_1\prec \dots \prec s_n$. If $\sm(L,n)<1$ (see Definition \ref{def:nakatsu}), then one can construct in $O(n)$ time an alphabetic code $C$ for which the codeword assigned to symbol $s_i$ has length upper bounded by  $\min(\ell_i, n-1)$, for each $i=1,\dots,n.$
\end{lemma}
}

\begin{algorithm}[H]\label{alg:ours}
\caption{BDDV algorithm}
Let $S=\{s_1,\dots,s_n\}$ be a set of symbols and $P=\mol{p_1,\dots,p_n}$ the associated probability distribution.

Construct the extended distribution with $2n-1$ elements 
$Q = \mol{q_1,\ldots,q_{2n-1}} = \mol{p_1,0,p_2,\dots, p_{n-1},0,p_n}$.

Compute the lengths $L=\mol{\ell_1,\dots,\ell_{2n-1}}$ as follows
\begin{equation*}
        \ell_i = \begin{cases}\label{defell}
            k & \text{ if } q_i=0,\\
            \lceil-\log q_i \rceil & \text{ if } i=1 \text{ or } i=2n-1,\\
            \lceil-\log q_i \rceil +1 & \text{ if } q_i > 0,
        \end{cases}
\end{equation*}
where $k = \max_{i} \lceil-\log p_i \rceil +1$.

{Apply Lemma \ref{lemma:linear_time_alg} on the lengths $L$ to construct an alphabetic code $C'$ for $Q$ such that the $i^{th}$ codeword has length at most $\min(\ell_i,2n-2)$ for each $i=1,\dots, 2n-1$ (for  details on the linear time procedure, see \cite{noi}).}

Prune the binary tree representing
$C'$ by removing the additional $n-1$ leaves corresponding to the 
$n-1$ zero probabilities in $Q$, 
in order to obtain an alphabetic code $C$ for $S$.    
\end{algorithm}
Let us observe that the 
correctness of the algorithm derives from the fact that the lengths $L$ defined in step 3 of Algorithm \ref{alg:ours} {satisfy the condition of Theorem \ref{th:nakatsu_condition} and Lemma \ref{lemma:linear_time_alg}. Therefore, an alphabetic code with lengths $L$ exists, and it can be constructed in linear time. We can summarize the results as follows.}

\begin{theorem}[\cite{noi}]\label{th:noi}
    For any set of symbols $S=\{s_1,\dots,s_n\}$, $s_1 \prec \dots \prec s_n$, with associated probabilities $P=(p_1,\dots,p_n)$, 
    one can construct in linear time an alphabetic code $C$ for $S$ whose average length satisfies
    \begin{align*}
        \mathbb{E}[C] &\leq H(P) + 2 - p_1 \left (2-\log p_1 -\lceil -\log p_1 \rceil\right)
        - p_n \left (2-\log p_n -\lceil -\log p_n \rceil\right)\\
        & \ \ \ - \sum_{i=1}^{n-1} \min(p_i,p_{i+1})\\
        &< H(P) + 2 - p_1 - p_n - \sum_{i=1}^{n-1} \min(p_i,p_{i+1}).
    \end{align*}
\end{theorem}
In \cite{noi} the authors also provided further improvements on particular 
probability distribution instances, as stated in the following theorem. 
We recall that a probability  distribution $P = \mol{p_1,\dots,p_n}$ is \textit{dyadic}
if each  $p_i$ is equal to $2^{-k_i}$, for suitable integers
$k_i>0$.
\begin{theorem}[\cite{noi}]
    For any dyadic distribution $P=\mol{p_1,\dots,p_n}$ on the set of symbols $S=\{s_1,\dots,s_n\}$, $s_1 \prec \dots \prec s_n$,  one can construct in linear time an alphabetic code $C$ for $S$ whose average length $\mathbb{E}[C]$ satisfies 
    \begin{equation*}
        \mathbb{E}[C] \leq H(P)+1 - p_1 -p_n.
    \end{equation*}
\end{theorem}

\section{Variations and generalizations}\label{vargen}
In this section, we will survey the known results 
about variations and generalizations of the classical problem 
of constructing alphabetic codes of minimum average length.

\subsection{Alphabetic codes optimum under different criteria}\label{diff}
In the classical formulation  of the problem, one is  given a sequence of positive weights 
$w=\mol{w_1,\dots,w_n}$, which usually is assumed to be a probability distribution, and  the objective is to construct an alphabetic code for $w$ of minimum average cost, that is, one would like to compute the 
quantity
\begin{equation}\label{media}
    \min \sum_{i=1}^n w_i \ell_i,
\end{equation}
where $\ell_i$ is the length of the codeword associated with the weight $w_i$. However, in some cases, it is more useful to consider other criteria for the construction of the optimal code. 

Hu, Kleitman and Tamaki in \cite{Hu_Kleitman} considered the variant
of the problem (\ref{media}) in which instead of minimizing the average length of the  tree
representing the alphabetic code, they want to minimize 
suitable cost functions that they call
\textit{regular cost functions}. As an example of a regular cost function in \cite{Hu_Kleitman}, they consider the following one: 
\begin{equation}\label{minmax}
    \max_i w_i 2^{\ell_i}.
\end{equation}
Alphabetic trees optimized 
according to (\ref{minmax}) are also called \textit{alphabetic minimax} trees.
Other examples of regular cost functions are 
described in \cite{Hu_Kleitman}, together with their justifications.
Moreover, the authors of \cite{Hu_Kleitman} observed that through a suitable modification, the Hu-Tucker algorithm can be used to compute an optimal alphabetic tree for any arbitrary regular cost function in $O(n \log n)$ time. Successively, for the specific case of alphabetic minimax trees, Kirkpatrick and Klawe \cite{kirkpatrick1985alphabetic} improved the result
given in \cite{Hu_Kleitman}. Indeed, they proposed a linear time algorithm for the problem when the weights are integers (actually, their algorithm minimizes
the quantity $\max_i \{w_i + \ell_i\}$, but one can see this 
minimization is equivalent to (\ref{minmax}), as discussed in \cite{gawrychowski2013alphabetic}). Later, Gagie \cite{gagie2009new} provided a $O(n d \log \log n)$ time algorithm
for the same problem considered in \cite{kirkpatrick1985alphabetic}, where $d$ is the number of distinct integers 
in the set $\{\lceil w_1\rceil,\ldots,\lceil w_n\rceil\}$.
A more efficient algorithm has been designed by Gawrychowski \cite{gawrychowski2013alphabetic}, 
who gave  a $O(nd)$ algorithm for the construction of the optimal alphabetic minimax tree, 
where $d$ is the number of distinct integers in the set $\{\lfloor w_i\rfloor,\ldots,\lfloor w_n\rfloor\}$.

\smallskip
In the paper \cite{Za} the author considered the following problem. Given a sequence of positive weights $w=\mol{w_1,\dots,w_n}$, and an alphabetic tree $T_n$ (i.e., a tree representing
an alphabetic code for $w$), one defines the cost $w(T_n)$ of $T_n$ in the following way:
\begin{itemize}
    \item the cost of the $i^{th}$ leaf (read from left to right) is equal to $w_i$;
    \item the cost $w(u)$ of any internal node $u$ of $T_n$ is given by
    \begin{equation*}
    w(u)=\max\{w(u_\ell), w(u_r)\},
    \end{equation*}
    where $u_\ell$ is the left child and $u_r$ is the right child of $u$, respectively;
    \item the cost $w(T_n)$ is 
      \begin{equation}\label{Z}
    w(T_n)=\sum w(u),
    \end{equation}
    where the summation is over all internal nodes of $T_n$.
\end{itemize}
One can see that the kind of minimization problem described above does \textit{not} 
fit in the framework of regular cost functions considered in \cite{Hu_Kleitman}.
The author of \cite{Za} gives a $O(n\log n)$ algorithm to compute an alphabetic tree whose  cost (as 
defined in (\ref{Z})) is minimum.

Fujiwara and Jacobs \cite{fujiwara2014huffman} analyzed a generalization of the classical 
alphabetic-tree problem, called \textit{general cost alphabetic} tree, where instead of associating a weight to each leaf, we associate an arbitrary function. More formally, the problem can be described as follows:
 Given $n$ arbitrary functions $f_1,\dots,f_n:\mathbb{N} \to \mathbb{R}_+$, the objective is to construct an alphabetic tree such that 
 \begin{equation}\label{eq:FJ}
     \sum_{i=1}^n f_i (\ell_i)
 \end{equation}
is minimized, where $\ell_i$ is the depth of the $i^{th}$ leaf from left to right. The authors of
\cite{fujiwara2014huffman}
showed that the dynamic programming approach for the classical alphabetic tree problem can be extended to arbitrary cost functions, obtaining a $O(n^4)$ time and $O(n^3)$ space algorithm for the construction of an optimal general cost alphabetic tree. In addition, they extended 
their results to Huffman codes with general costs. However,  unlike the case of alphabetical trees, in the Huffman scenario,  they showed that the general problem becomes NP-hard.

Given a sequence of positive weights $w=\mol{ w_1,\dots,w_n}$, Baer \cite{baer2} considered the  
problem of minimizing   the function
$$\log_a\left (\sum_{i=1}^nw_ia^{\ell_i}\right),$$ 
 for a given $a\in (0,1)$, over the class of all alphabetic codes with $n$ codewords.
The author gave  a $O(n^3)$ time and $O(n^2)$ space dynamic programming algorithm 
to find the optimal solution, and claimed that 
  methods traditionally used to improve the speed of optimizations in related problems, such as the Hu–Tucker procedure, fail for this problem. 
  In the same paper, Baer  introduced two algorithms that can find a suboptimal solution in linear time (for one) or $O(n\log n)$
 time (for the other), and provided redundancy bounds guaranteeing their coding efficiency.


\subsection{Height-limited alphabetic trees}\label{sec:limited}
In contexts of routing lookups \cite{gupta2000near,Nagaraj2011482} it emerges the
necessity
to minimize
the average packet lookup time while keeping the worst-case lookup time \textit{within} 
a fixed bound.
Such a need directly translates into considering a specific class of alphabetic trees known as
\textit{height-limited} alphabetic trees. In this scenario, the main problem is to find
alphabetic codes of minimum average length, under the constraint that no word in the code has
a length above a certain input parameter.
More formally, 
given an ordered set of symbols 
$S=\{s_1, \ldots , s_n\}$,  a 
 probability distribution $P=\mol{p_1, \ldots , p_n}$ on $S$, 
 one  seeks to solve the following optimization problem:
 \begin{align*}
\min_{C \mbox{ \footnotesize alphabetic}}\mathbb{E}[C]&= 
\min_{C \mbox{ \footnotesize alphabetic}}\sum_{i=1}^m p_i\,\ell_i,\\
\mbox{subj. to } \ell_i \leq L, &\quad \forall i=1,\dots,n,
 \end{align*}
where $\ell_i$ is the length of the codeword associated to the symbol $s_i$ of 
probability $p_i$. Such codes are also known as $L$-restricted alphabetic codes.

In \cite{hu1972path}, Hu and Tan presented an algorithm for constructing an optimal binary tree with the
restriction that its height cannot exceed a given integer $L$. However, the time complexity of their algorithm is exponential in $L$.
Garey \cite{garey1974optimal} gave an $O(n^3 \log n)$ time algorithm for the construction of an optimal height-limited alphabetic tree. Successively, Itai \cite{itai} and Wessner \cite{wessner1976optimal} independently reduced this time to $O(n^2 L)$. Hassin and Henig \cite{HASSIN1993221} extended a monotonicity theorem of Knuth \cite{knuthp} to hold under weaker assumptions and applied this new result to reduce the complexity of several 
optimization scenarios, 
including height-limited alphabetic trees. Similarly, Larmore \cite{Larmore} designed
a different  $O(n^2 L)$ time algorithm for the construction of an optimal height-limited alphabetic tree by improving Hu and Tan's algorithm \cite{hu1972path}. 
Successively, Larmore and Przytycka \cite{Larmore_Prz} provided an $O(n L \log n)$ time algorithm for the construction of an optimal alphabetic tree with height restricted to $L$. 

Gupta \textit{et al.} \cite{gupta2000near} focused their attention on the construction of nearly optimal $L$-restricted alphabetic codes via an $O(n \log n)$ time algorithm for constructing an alphabetic tree whose average length differs from the optimal value by at most 2. Similarly, Laber \textit{et al.} \cite{Laber} suggested a simple approach to construct sub-optimal $L$-restricted alphabetic codes, comparing their average length with the average length of the Huffman code.

\subsection{Binary trees that are alphabetic with respect to 
given partial orders}\label{partord}
In the classic alphabetic tree problem, there is a total order relation $\prec$ on the set of symbols $S=\{s_1,\dots,s_n\}$, and the left-to-right reading of the tree leaves
must give the same ordering of the elements in $S$, according to the relation $\prec$.

However, in some situations (see, e.g., \cite{Lipman})
there might be given only a \textit{partial} order on the set of 
symbols $S=\{s_1,\dots,s_n\}$, and the problem is to construct a tree (i.e., a code)
in which the left-to-right reading of the leaves of the tree is \textit{consistent} 
with the partial ordering on $S$.
Lipman and Abrahams \cite{Lipman} studied such a variant of the basic problem. They were 
motivated by questions that arise when one wants to 
detect defects in a pipeline. 
The authors of \cite{Lipman} proposed to solve the problem indirectly, that is, by employing
the idea of decomposing the given partial order into a set of linear total orders.
Subsequently, Lipman and Abrahams applied the  Hu-Tucker algorithm \cite{hu1971optimal} to each subproblem (corresponding to
each linear order obtained from the decomposition) and constructed 
  the final tree by applying,  again,  the Hu-Tucker algorithm to the several 
partial solutions previously obtained. In \cite{Lipman} the authors left open the problem of providing explicit upper bounds on the average length of the trees produced by their construction.

Later,  Barkan and Kaplan \cite{barkan2002partial} studied a problem similar to the one considered 
by Lipman and Abrahams \cite{Lipman}.
In particular, Barkan and Kaplan \cite{barkan2002partial}   
addressed a generalized version of the classic problem, that they called \textit{partial alphabetic} tree problem. 
In the partial alphabetic tree problem one is given a multiset of non-negative weights $W=\{w_1,\dots,w_n\}$, partitioned into $m\leq n$ blocks $B_1,\dots,B_m$. The objective is
to build  a tree $T$,  where the elements of $W$ reside in its leaves, 
satisfying the following property:  If we traverse the leaves of $T$
from left to right, then all leaves of $B_i$ precede all leaves of $B_j$ for every $i < j$. Furthermore, among all such trees, 
it is required that $T$ has minimum average length
\begin{equation*}
    \sum_{i=1}^n w_i \ell_i,
\end{equation*}
where $\ell_i$ is the depth of $w_i$ in $T$. In \cite{barkan2002partial} the authors designed a pseudo-polynomial time algorithm for the construction of an optimal tree,
whose complexity depends on the weight values. Moreover, the  technique 
developed in \cite{barkan2002partial} is general enough 
to apply to several other objective functions,
possibly different from the average length of the tree. 
However, the problem of whether there exists or not 
an algorithm for the partial alphabetic tree problem that runs in \textit{polynomial time},
for any set of weights,
is still an open question.

\subsection{Alphabetic AIFV codes}\label{AIFV}
Prefix codes, a superset of alphabetic codes, are an example of uniquely decodable codes that 
allow \textit{instantaneous} decoding. Instantaneous decoding refers to the following
important property:   Each codeword in any string of codewords can be 
uniquely decoded
(reading from left to right) as soon as it is received.

Yamamoto \textit{et al.} \cite{Yamamoto} 
introduced binary AIFV (Almost Instantaneous Fixed-to-Variable length) codes as 
 uniquely decodable codes that might suffer 
of at most two-bit decoding delay, that is,
for which each codeword in any string of codewords can be 
uniquely decoded
 as soon as its bits, plus two more,  are received.
The authors of \cite{Yamamoto} showed that
the use  of AIFV codes can improve
the compression of stationary memoryless sources, in some circumstances.
Successively, Hiraoka and Yamamoto \cite{HiYa} defined the alphabetic version of the AIFV codes by using three code trees for the decoding process with at most a two-bit decoding delay. They also proposed an algorithm for the construction of almost optimal binary alphabetic AIFV codes by modifying Hu-Tucker codes \cite{hu1971optimal}. However, despite their non-optimality, the constructed codes still attain a better compression rate than
classical Hu-Tucker codes. Later, Iwata and Yamamoto \cite{IY} proposed a natural extension of the alphabetic AIFV codes, called alphabetic AIFV-$m$ codes.
Such codes use $2m-1$ code trees in the decoding phase with at most $m$-bit decoding delay for any integer $m\geq 2$. The authors also designed a polynomial time algorithm for the construction of an optimal binary AIFV-$m$ code.
 
\subsection{Linear time algorithms for special cases}
In Section \ref{sec:optimal} we have 
mentioned that the best-known algorithms for the construction of optimal alphabetic codes have
a $O(n \log n)$ time complexity in the general case. However, 
under special circumstances, it is possible to 
obtain linear algorithms for the problem. 

Klawe and Mumey \cite{Klawe}
extended the ideas and techniques of Hu and Tucker 
and designed a $O(n)$-time algorithm 
to construct optimal alphabetic codes 
either when  all the input weights are within a constant factor of one another or
when they are exponentially separated. A sequence $w_1,\dots, w_n$ of weights is said to be \textit{exponentially separated} if there exists a constant $C$ such that it holds that
$$
    |\{i: \lfloor \log w_i \rfloor = k\}| < C, \forall k \in \mathbb{Z}.
$$
Subsequently, Larmore and Przytycka \cite{LARMORE19981} considered the \textit{integer alphabetic} tree problem, where the weights are integers in the range $[0,n^{O(1)}]$. The authors provided a $o(n\log n)$ algorithm for the construction of an optimal integer alphabetic tree. Moreover, by relating the complexity of the optimal alphabetic tree problem to the complexity of sorting, they gave an $O(n \sqrt{\log n})$-time algorithm for the cases in which the weights can be sorted in linear time, or equivalently the weights are all integers in a small range \cite{Hu_int}. Successively, Hu \textit{et al.} \cite{Hu_int} further improved the results of \cite{LARMORE19981} by designing an $O(n)$-time algorithm for the construction of an optimal integer alphabetic tree. 
In \cite{Hu_Morgen}, Hu and Morgenthaler analyzed several classes of inputs on which the Hu-Tucker algorithm \cite{hu1971optimal} runs in linear time. For example, they showed that for \textit{almost uniform}
sequences of weights, that are sequences $W$ of weights for which
\begin{equation*}
    \forall w_i,w_j,w_k \in W, \quad w_i + w_j \geq w_k,
\end{equation*}
and also for bi-monotonal increasing sequences, i.e.,  sequences of weights for which
\begin{equation*}
    w_1 + w_2 \leq w_2 + w_3 \leq \dots \leq w_{n-1}+w_n,
\end{equation*}
holds, 
the Hu-Tucker algorithm requires only linear time.

In \cite{hu1982combinatorial}, Hu introduced the notion of \textit{valley sequence}
to the purpose of understanding the computational
complexity of constructing optimal
alphabetic codes. 
We recall that a sequence $w_1,\dots,w_n$ of weights is a valley sequence if 
$$
    w_1 > w_2 >\dots >w_{j-1}\leq w_j \leq w_{j+1}\leq \dots \leq w_n.
$$
In other words, the weights are first decreasing and then increasing.
Hu \cite{hu1982combinatorial} showed that if the weight sequence $W$ is a valley sequence, then the cost of the optimal alphabetic tree for $W$ is the same as the cost of the Huffman tree for $W$. In addition, Hu proved
that one can construct an optimal alphabetic tree for a valley sequence in linear time. 
Moreover, since an ordered sequence is just a special case of a valley sequence 
one can also construct an optimal alphabetic tree for an ordered sequence in linear time. 
A similar result for ordered sequences also derives from the well-known fact that 
the minimum average length of an alphabetic code for an ordered sequence $W$, is equal to 
the minimum average length of a prefix code for $W$ \cite{yeung1991alphabetic}. 
Therefore, since Huffman codes for ordered sequences can be computed in linear time \cite{vL}, one gets that 
minimum average length alphabetic codes for ordered sequences can also be computed in linear time.

\subsection{$k$-ary alphabetic trees}
Most of the literature on alphabetic codes concerns \textit{binary} alphabetic codes.
In this section, we will describe the few known results about general 
$k$-ary alphabetic codes, for arbitrary $k\geq 2$.
The papers containing results on $k$-ary alphabetic codes use the terminology of search trees. Since we have already
seen that there is an equivalence between alphabetic codes and search algorithms 
(search trees)  
that operate through comparison 
tests, in this section we stick to the search-tree terminology.

The first author to study the problem of constructing optimal (i.e., minimum 
average length) $k$-ary alphabetic trees was Itai, in  \cite{itai}. In that paper, 
the author claimed  a $O(n^2L\log k)$ time algorithm for constructing optimal 
$k$-ary alphabetic trees of maximum depth $L$, which is the
same scenario considered in Section \ref{sec:limited}, and a 
$O(n^2\log k)$ time algorithm for constructing \textit{unrestricted}  optimal 
$k$-ary alphabetic trees. Subsequently,  Gotlieb and  Wood \cite{Go} pointed out a gap
in the analysis of \cite{itai}, invalidating its claims. Moreover,
the authors of \cite{Go} designed a $O(n^3L\log k)$ time algorithm for constructing optimal 
$k$-ary alphabetic trees of maximum depth $L$  and a 
$O(n^3\log k)$ time algorithm for constructing \textit{unrestricted}  optimal 
$k$-ary alphabetic trees. 
Ben-Gal \cite{ben2004upper} considered the problem of constructing \textit{almost} 
optimal $k$-ary alphabetic trees. In particular, he generalized the  weight-balancing
algorithm by Horibe \cite{HORIBE1977148} (that we have described in Section \ref{sec:upp}) to the
arbitrary case of $k\geq 2$, and provided some upper bounds
on the average length of the $k$-ary alphabetic trees one obtains by
applying his method.
Kirkpatrick and Klawe \cite{kirkpatrick1985alphabetic} considered the $k$-ary alphabetic minimax tree problem, which is a generalization of the problem discussed in Section \ref{diff}, providing a linear time algorithm when the weights are integers and a $O(n \log n)$ time algorithm for the general case. Subsequently, Coppersmith \textit{et al.} \cite{Coppersmith} considered a variant of the problem in \cite{kirkpatrick1985alphabetic}, in which each internal node of the tree has degree at most $k$ (not exactly $k$ as in \cite{kirkpatrick1985alphabetic}). They gave a linear-time algorithm, when the input weights are integers, and an $O(n \log n)$ time algorithm for real weights. Moreover, they provided a tight upper bound for the cost of the constructed solution. Gagie \cite{gagie2009new} developed a $O(n d \log \log n)$ time algorithm for the same problem considered in \cite{kirkpatrick1985alphabetic}, where $d$ is the number of distinct integers in the set $\{\lceil w_1\rceil,\ldots,\lceil w_n\rceil\}$. The algorithm improves upon the previous result of \cite{kirkpatrick1985alphabetic} when $d$ is small.

\section{Miscellanea}\label{misc}

In this section, we will review a few interesting
results on alphabetic codes that deal with disparate
problems, not classifiable under a unified theme.

Let  $S=\{s_1, \ldots , s_n\}$ be  an  ordered set of symbols 
and $P=\mol{p_1, \ldots , p_n}$ be the ordered
probabilities of the symbols in $S$.
From the definition, one has that the minimum average length of 
any alphabetic encoding of
$S$, regarded as a function of $P$, is \textit{not} invariant 
with respect to permutations of $p_1, \ldots , p_n$.
By  contrast, if $S$ is unordered then one has 
that the minimum average length of a prefix encoding of
$S$ (i.e., the average length of a Huffman code for $S$), \textit{is invariant} 
with respect to permutations of $p_1, \ldots , p_n$.
Therefore, the following problem naturally arises: Given the ordered set of symbols 
$S=\{s_1, \ldots , s_n\}$,
and an arbitrary probability distribution  $p_1, \ldots , p_n$, what is the
ordering of $p_1, \ldots , p_n$ that forces the average length of an
optimal alphabetic code for $S$ to assume its \textit{maximum} value?
This problem has been studied by Kleitman and Saks~\cite{ks} who proved the following
neat result. Given $P=\mol{p_1, \ldots , p_n}$, such that $p_1\leq p_2\leq \ldots , \leq p_n$,
then the permutation of the elements of $P$ that produces the costliest
optimal alphabetic code is given by $p_1, p_n, p_2, p_{n-1}, \ldots, $.
Moreover, the authors of \cite{ks} expressed the average length of such 
costliest 
optimal alphabetic code in terms of the average length of a Huffman code
for a probability distribution $Q$, easily computable from $P$. 
The papers \cite{hu1972least,ric} considered strictly related problems, that
can be derived as corollaries of the main result of \cite{ks}. Finally Yung-chin
\cite{yung1985extended} extended the main result of \cite{ks} to the
case of alphabetic codes with a hard limit on the maximum codeword
length, that is, in the same framework considered in Section \ref{sec:limited}.

In the paper \cite{RAMANAN1992279}, Ramanan considered the important problem
of efficiently testing whether or not a given alphabetic code is optimal 
for an ordered set of symbols $S=\{s_1, \ldots , s_n\}$,
and the associated probability distribution  $P$.
Using the proof of correctness of the Hu-Tucker algorithm \cite{hu1971optimal}, 
the author of \cite{RAMANAN1992279} gives
 necessary and sufficient conditions on the sequence $P=\mol{p_1, \ldots, p_n}$,
 for a given code tree to be optimal.  From this result,
 Ramanan shows that the optimality of very skewed trees (i.e. trees in which the number of nodes in each level is bounded by some constant) can be tested in linear time. 
 Ramanan also shows that the optimality of well-balanced trees (i.e. trees in which the maximum difference between the levels of any two leaves is bounded by some constant) can also be tested in linear time. The general
 case of testing the optimality of an \textit{arbitrary} code  tree  in linear time is left
 open, and it represents one of the main open problems in the area.

In the paper \cite{anily1989ranking}, Anily and Hassin considered the problem of ranking the best 
$K$ trees, given   weights $w_1, \ldots , w_n$. More precisely, 
given weights $w_1, \ldots , w_n$, the problem is that of computing 
the binary tree with the smallest average length, the binary tree with the second
smallest average length, ..., the binary tree with the $K^{th}$ smallest average length.
The authors studied both the alphabetical and non-alphabetical cases. 
In particular, they presented an $O(Kn^3)$ time algorithm for ranking both the $K$-best binary alphabetic trees and the $K$-best binary non-alphabetic trees. 

In \cite{larmore1989minimum}, Larmore introduced the concept of \textit{minimum delay} codes. A minimum delay code is a prefix code in which, instead of minimizing the average length, the aim is to minimize the \textit{expected delay} of the code, which is the expected time between a request to transmit the symbol and the completion of that transmission, assuming a channel with fixed capacity, where requests are queued. Larmore formally defined the expected delay as a particular nonlinear function of the average code length and gave an $O(n^5)$ time and $O(n^3)$ space algorithm to find a prefix code of minimum expected delay. Moreover, the algorithm can be also adapted to find an alphabetic code of minimum expected delay. However, there is no guarantee that the algorithm will require polynomial time since its complexity strongly depends on the monotonicity of the weights. Therefore, the question of whether a polynomial-time algorithm exists for constructing an alphabetic code of minimum delay remains open.

In \cite{julia0}, Abrahams considered the problem of constructing codes with 
monotonic
codeword lengths (monotonic codes), and 
optimal monotonic codes were examined in comparison with optimal
alphabetic codes. 
Bounds between their lengths were  derived, and sufficient conditions were 
given such that the Hu-Tucker algorithm can be used to find the optimal monotonic code.

In \cite{julia1}, Abrahams considered 
 a parallelized version of the search problem described in Section \ref{AlpSear}. 
 More precisely, 
a  natural parallelized version of the classical search problem is to
distribute the set of $n$ items into $k$ subsets, each one of which is to be
searched simultaneously (i.e., in parallel)  for the single item of interest.
The case that $k = 1$ corresponds exactly to the classical case of alphabetic codes. 
The case 
$k = n$ is trivial: one item
is placed in each subset. It is of interest to resolve the intermediate
cases: which of the items, occurring with probabilities $p_1, \ldots , p_n$,
should be placed into a common subset, to be searched by the
 Hu-Tucker algorithm respectively within that subset, so
as to minimize average search length? 
In \cite{julia1}, Abrahams
gave an algorithm for this problem and provided upper bounds 
 on the resulting
minimum average search time.

In \cite{baer1}, Baer constructs   alphabetic 
codes  optimized for  power law distributions, that is, when the
probability of the $i^{th}$ symbol $p_i$ is of the form $p_i\sim ci^{-\alpha}$,
where $c$ and $\alpha>1$ are constant, and $f(i)\sim g(i)$
means that the ratio of the two functions goes to 1 with increasing $i.$

In \cite{bruno2025optimal}, Bruno \textit{et al.} studied the problem of designing optimal binary prefix and alphabetic codes under the constraint that each codeword contains at most $D$ ones. They proposed an $O(n^2D)$-time dynamic programming algorithm for the construction of such optimal codes, and derived a Kraft-like inequality for their existence.

In \cite{ko} Kosaraju \textit{et al.} introduced the \textit{Optimal Split Tree} problem. Let $A = \{a_1,\dots,a_n\}$ be a set of elements where each element $a_i$ has an associated weight $w_i > 0$. A partition of $A$ into two subsets $B, B\setminus A$ is called a \textit{split} of $A$. A set $S$ of splits of $A$ is a \textit{complete set of splits} if for each pair $a_i,a_j \in A$ there exists a split $B, B \setminus A$ in $S$ such that $a_i \in B$ and $a_j \in B \setminus A$. A \textit{split tree} for a set $A$ and a set $S$ of splits of $A$ is a binary tree in which the leaves are labeled with the elements of $A$ and the internal nodes correspond to the splits in $S$. More formally, for any node $v$ of a binary tree, let $L(v)$ be the set of labels of the leaves of the subtree rooted at $v$. A split tree is a full binary tree such that for any internal node $v$ with children $v_1, v_2$ there exists a split $\{B_1,B_2\} \in S$ such that $B_1 \cap L(v) = L(v_1)$ and $B_2 \cap L(v) = L(v_2)$. Hence, given a set $A$ with its associated weights and a complete set of splits of $A$ the optimal split tree problem is to compute a split tree with minimum average length. One can see that the problem is a generalization of many others, including the Huffman tree problem and the optimal alphabetic tree problem. Indeed, when the set $S$ of splits contains all possible splits of $A$ we get the classic Huffman tree problem, while when the set $S$ of splits contains $\{B_1, A \setminus B_1\}, \dots, \{B_{n-1}, A \setminus B_{n-1}\}$ splits where $B_i = \{a \in A : a \prec a_i\}$, the problem reduces to the classic alphabetic tree problem. In \cite{ko}, Kosaraju \textit{et al.} showed that the optimal split tree problem, in the general case, is NP-complete. An equivalent proof of NP-completeness was previously provided by Laurent and Rivest \cite{laurent1976constructing}. Moreover, in \cite{ko} the authors gave an $O(\log n)$ approximation algorithm for the problem, providing also an example for which the algorithm achieves an $\Omega(\log n /    \log \log n)$ approximation ratio. In addition, they adapted their algorithm obtaining an $O(1)$ approximation algorithm for the partially ordered alphabetic tree problem (the same problem considered in Section \ref{partord}).  

In \cite{Le}, Levcopoulos \textit{et al.} prove the
interesting result that for any
arbitrarily small $\epsilon >0$, one
can construct, in time  $O(n)$,  an alphabetic tree whose
cost is within a factor of $(1+\epsilon)$ from the optimum.

\section{Open Problems}\label{open}
In this section, we list a few open problems in the area of alphabetic codes.
\begin{itemize}
\item 
In Section \ref{sec:optimal} we presented the Hu-Tucker
algorithm \cite{hu1971optimal} for
constructing alphabetic codes of minimum average length. The algorithm has time
complexity $O(n\log n)$, where $n$ is the number of components
in the input probability distribution $P$. To date, there is no non-trivial lower bound 
on the complexity of algorithms that construct alphabetic codes of minimum average length.
Therefore, the following question is still open: Are there algorithms for
constructing alphabetic codes of minimum average length with time complexity $o(n\log n)$?
\item A strictly related problem is the following:
Given a code $C$ and a probability distribution
$P=\mol{p_1, \ldots , p_n}$ can one check in time $O(n)$
whether or not $C$ is a minimal average length alphabetic code for $P$?
Here, and in the previously stated problem,
the computational complexity of the algorithm is measured in the
linear decision model (that, roughly speaking,  computes the complexity
of an algorithm based on the number of comparisons the algorithm performs to
produce the desired output). 
Some partial results are contained in \cite{RAMANAN1992279}, but the general problem
is still unresolved. It is interesting to note that the same problem for Huffman codes 
can be solved in time $O(n)$ \cite{RAMANAN1992279}.
One of the reasons that make the checking problem interesting, is that its computational
complexity is a lower bound on the complexity of \textit{finding}
an optimal alphabetic code.
\item Ambainis \textit{et al.} \cite{Am+} and Ahlswede and Cai \cite{AC}
consider a variant of comparison-based search algorithms, in which
the test outcome is received after $d$ time units the test has been performed, 
for some integer $d$. This implies
that the search algorithm has to execute the generic  $i^{th}$ query 
based on the knowledge of the answers to the queries from the $1^{st}$ to the $(i-d-1)^{th}$, 
\textit{only}.
This kind of algorithm generates alphabetic codes with a different structure
with respect to the classical ones. The papers quoted above study algorithms with optimal
\textit{worst-case} performances. It would be interesting to study alphabetic
codes arising from the same kind of algorithms but with optimal
\textit{average-case} performances.
\item Ahlswede and Cai \cite{AC1} studied the important problem of \textit{source identification}.
The basic setting is the following: one is given a set of source symbols $S$,  an encoding
$\co:S\mapsto \{0,1\}^+$, and an arbitrary $s\in S$. For any $s'\in S$, one wants to establish
whether it is the case, or not, that $s'=s$, and this identification process is performed 
by confronting, from left to right, whether or not the bits of $\co(s')$ coincides with
the bits of $\co(s).$ 
In their study, Ahlswede and Cai \cite{AC1} considered only prefix encoding.
However, they suggested to extend their findings to other classes of variable
length codes. To the best of our knowledge, the problem 
of source identification for alphabetic codes has never been studied, and it surely it
deserves to be.
\item In Section \ref{sec:upp} we have presented several results that upper bound
the average length of optimal alphabetic codes in terms of the entropy $H(P)$ plus some easy-to-compute function $f(P)$. Remarkably, there seems to be just one known lower bound on the 
 average length of optimal alphabetic codes of the form $H(P)+g(P)$, for suitable 
 functions $g(\cdot)$ (see  \cite{yeung1991alphabetic}).
 Due to the more strict constraints that 
 the alphabetic 
 codes have to satisfy, with respect to the Huffman codes,
 it seems that there could be room for improvement.
 Notice that several bounds of the form $H(P)+g(P)$ are known for
 binary search trees, e.g.~\cite{AW,aigner,DD96lowerbounds}, for which the
 knowledge of the
 probabilities of the internal nodes (i.e., of the successful searches) allow 
 to improve the bound
 that exploits only $H(P)$, e.g.,~\cite{DD96lowerbounds,M75}.
 
 \item In the paper \cite{dagan1}
 the authors extend the basic setting of alphabetic codes (equivalently, of comparison-based search procedures) to the case in which 
 ``lies" are present, that is, the case
 of search procedures that successfully determine the unknown element 
even though a fixed number $k$ of queries can receive an erroneous answer.
The authors of \cite{dagan1} prove the following strong result: there exists an algorithm
for the problem described above performing  an average number of queries upper bounded by
\begin{equation}\label{dagan1upp}
    H(P)+k\sum_{i=1}^np_i\log \log\frac{1}{p_i} + O\left (k\sum_{i=1}^np_i\log \log \log\frac{1}{p_i}
    +k\log k\right ).
\end{equation}
Moreover, \textit{any} such algorithms must perform an average number of queries lower
bounded by
\begin{equation}\label{dagan1low}
    H(P)+k\sum_{i=1}^np_i\log \log\frac{1}{p_i} - (k \log k + k + 1).
\end{equation}
It is only natural to ask whether one can tighten the gap between the upper bound (\ref{dagan1upp})
and the lower bound (\ref{dagan1low}).

\smallskip
\item In Subsection \ref{diff} we have described different results related to alphabetic codes 
optimum under different criteria. For some of these cases, it is known how to efficiently
compute an optimal alphabetic code. However, the problem of providing good upper
bounds on 
the cost of optimal solutions (as done in Section \ref{sec:upp} for the classic case) is wide open.

\smallskip
\item In Subsection \ref{partord} we have presented the known results on the area of alphabetic
codes for partially ordered sets. This a generalization of the classical problem,
in which
the mapping $\co:S=\{s_1, \ldots , s_n\}\mapsto \{0,1\}^+$ must preserve a given \textit{total} 
order on $S$, to the general case in which is given a partial \textit{order} relation
on $S$. Although some interesting results are known, 
 this research area is mostly unexplored.

 \smallskip
 \item In many applications, it is important to have variable length codes that satisfy properties
 stronger than the classical prefix property. One of these properties is the \textit{fix-free}
 property (see \cite{deppe} and the references therein quoted). Fix-free codes have the characteristic that no codeword is neither a prefix nor a suffix of any other in the codeset. 

 Another strengthening of the basic property of alphabetic codes would be to require
that they satisfy ``synchronizing properties", in the sense of \cite{cap2,cap1}.
It would be interesting to extend the known results about prefix alphabetic codes to fix-free alphabetic codes and synchronizing codes.

  \smallskip

  \item At the best of our knowledge, there are no algorithms that efficiently
  update the structure of
  optimal alphabetic codes, as the symbol probabilities change.
  It would be interesting to design such algorithms, in the same spirit of
  what has been done by Knuth \cite{knuthdin} for classical Huffman codes. 

  \smallskip
  
  \item Given a probability distribution $P=\mol{p_1, \ldots , p_n}$, there can be 
  {\textit{several}} different alphabetic codes having a minimum average length.
  In this case, it would be interesting to modify the known algorithms for 
  constructing minimum average length alphabetic codes that have the additional
  property of minimizing the maximum length, or other parameters of interest. For Huffman
  codes, this problem has been studied in \cite{sc}.

    \smallskip
 \item 
Finally, in \cite{depriscopersiano,Le+}, the authors have considered binary
alphabetic codes in which each codeword must satisfy given constraints
on the number of zeroes and ones it can contain. It would be interesting
to design efficient algorithms for
constructing minimum average-length alphabetic codes
in this scenario.

\end{itemize}

\end{document}